\documentclass[%
reprint,
superscriptaddress,
 amsmath,amssymb,
 aps,
prb,
]{revtex4-2}
\usepackage{csquotes}
\usepackage{relsize}
\usepackage{graphicx}% Include figure files
\usepackage{dcolumn}% Align table columns on decimal point
\usepackage{graphicx}
\usepackage{bm}% bold math
\usepackage[colorlinks=true,linkcolor=blue,citecolor=blue,filecolor=blue,urlcolor=blue]{hyperref}% add hypertext capabilities
\usepackage{cleveref}
\begin{document}

\title{Second-harmonic generation in plasmonic waveguides \\with nonlocal response and electron spill-out}% Force line breaks with \\

\author{Ahsan Noor}
\affiliation{Istituto Italiano di Tecnologia, Center for Biomolecular Nanotechnologies,\\ Via Barsanti 14, 73010 Arnesano, Italy.}
\affiliation{Dipartimento di Ingegneria Elettrica e dell'Informazione, Politecnico di Bari,\\ Via Re David 200, 70125 Bari, Italy.}
\author{Muhammad Khalid}
\affiliation{Istituto Italiano di Tecnologia, Center for Biomolecular Nanotechnologies,\\ Via Barsanti 14, 73010 Arnesano, Italy.}
\author{Federico De Luca}
\affiliation{Istituto Italiano di Tecnologia, Center for Biomolecular Nanotechnologies,\\ Via Barsanti 14, 73010 Arnesano, Italy.}
\affiliation{Dipartimento di Matematica e Fisica ``E. De Giorgi'', Universit\`a del Salento,\\ Via Arnesano, 73100 Lecce, Italy.}
\author{Henrikh M. Baghramyan}
\affiliation{Istituto Italiano di Tecnologia, Center for Biomolecular Nanotechnologies,\\ Via Barsanti 14, 73010 Arnesano, Italy.}
\author{Michele Castriotta}
\affiliation{Istituto Italiano di Tecnologia, Center for Biomolecular Nanotechnologies,\\ Via Barsanti 14, 73010 Arnesano, Italy.}
\affiliation{The Open University Affiliated Research Centre at Istituto Italiano di Tecnologia (ARC@IIT), \\ Via Morego 30, 16163 Genova, Italy.}
\author{Antonella D'Orazio}%
\affiliation{Dipartimento di Ingegneria Elettrica e dell'Informazione, Politecnico di Bari,\\ Via Re David 200, 70125 Bari, Italy.}
\author{Cristian Cirac\`i}
\email{cristian.ciraci@iit.it}
\affiliation{Istituto Italiano di Tecnologia, Center for Biomolecular Nanotechnologies,\\ Via Barsanti 14, 73010 Arnesano, Italy.}
\date{\today}% It is always \today, today,
\begin{abstract}
Plasmonic waveguides provide an integrated platform to develop efficient nanoscale ultrafast photonic devices.
Theoretical models that describe nonlinear optical phenomena in plasmonic waveguides, usually, only incorporate bulk nonlinearities, while nonlinearities that arise from metallic constituents remained unexplored.
In this work, we present a method that enables a generalized treatment of the nonlinearities present in plasmonic waveguides and use it to calculate second-harmonic generation from free electrons through a hydrodynamic nonlocal description.
As a general application of our method we also consider nonlinearities arising from the quantum hydrodynamic theory with electron spill-out.
Our results may find applicability in design and analysis of integrated photonic platforms for nonlinear optics incorporating wide variety of nonlinear materials such as heavily doped semiconductors for mid-infrared applications.
\end{abstract}

\maketitle

%\tableofcontents

\section{\label{sec:level1}Introduction}
Plasmonic systems provide the possibility of concentrating and manipulating light below the diffraction limit and are at the core of a variety of optical applications \cite{Stockman_2018,stockman2011}, from improved chemical and biological sensing \cite{anker2010biosensing,PhysRevLett.83.4357}, and efficient photovoltaic energy harvesting \cite{atwater2011plasmonics}, to ultrafast photonic signal processing \cite{macdonald2009ultrafast, HabibZhuFongYanik+2020+3805+3829}, and nanolasing \cite{johnson2002single,duan2003single,oulton2009plasmon}.
In the past decades, due to the ever-increasing demand for data processing capabilities, researchers have focused a great effort into the development of ultra-compact photonic elements, including plasmonic components, such as waveguides and couplers, \cite{ZIA200620,yang2012subwavelength,Han_2012,fang2015nanoplasmonic}, digital gates \cite{wei2011quantum,wei2011cascaded}, routers \cite{chang2007single,fang2010branched}, photon-electric converters \cite{shalin2014nano}, and control switches \cite{ming2010resonance}.
Plasmonic waveguides have also been relevant with regards to several quantum optical phenomena like single photon emission \cite{PhysRevLett.97.053002,Huck}, energy transfer and superradiance of emitter pairs \cite{martin2010resonance}, and qubit-qubit entanglement generation \cite{PhysRevLett.106.020501}.

Plasmonic systems allow miniaturization below the diffraction limits thanks to surface plasmon-polariton (SPP) modes \textemdash the resonant collective oscillations of free electrons (FEs) \textemdash appearing in materials with a high carrier concentration (i.e., metals and heavily doped semiconductors) and arising at the interface with a dielectric because of the interaction with an external electromagnetic (EM) excitation.
Localization of light associated to SPPs modes is naturally promising for the enhancement of intensity-dependent phenomena \cite{klein2006second,Zeng:2009hd,schuller2010plasmonics,Scalora:2010kd,Ciraci:2012vw,kauranen2012nonlinear,krasavin2018free,bonacina2020harmonic,tuniz2021nanoscale, C5CS00050E,DeLuca:19}.

Functionalities based on nonlinear optics are very attractive in terms of their femto-second response times and terahertz bandwidths. However, sizeable nonlinear effects demand both high field intensities and large interaction volumes, together with configurations that offer efficient nonlinear conversions as well as materials with large nonlinear susceptibilities \cite{Boyd:2006uq,Garmire:13,boardman2012advanced}.
All these features could be in principle provided by plasmonic systems, since metals possess some of the largest nonlinear susceptibilities.
Notably, however, interaction volumes in nanoantennas are quite limited and nonlinear efficiencies remain overall very small \cite{klein2006second,Zeng:2009hd,Scalora:2010kd,Ciraci:2012vw}.
On the other hand, plasmonic waveguides can sustain sub-wavelength field localization for the entire propagation length, thereby providing ideally larger volumes of interactions. 
Indeed, hybrid dielectric-plasmonic waveguides have been reported with a variety of nonlinear applications (see for example a comprehensive review on latest advances in nonlinear plasmonic waveguides \cite{tuniz2021nanoscale}). 
Most waveguide systems can be easily studied by decoupling the propagation and transverse problems \cite{Ruan:09,davoyan2010backward,Zhang:13,Zhang:13a,Wu:14,Sun:15,Huang:16,shi2019efficient}.
This separation is only possible when the electric field divergence, which is non-zero at the metal surface, is negligible.
As it will be shown in Sec. II, such approximation does not hold when nonlinearities arise directly from the plasmonic material \cite{Thyagarajan:12} and, in particular, from the dynamics of non-equilibrium FEs \cite{Ginzburg_2013,kauranen2012nonlinear,krasavin2018free}.
Indeed, FE nonlinearities in noble metals have been shown to strongly contribute to second-order nonlinear processes  in the visible/near-infrared (IR) \cite{Scalora:2010kd,Ciraci:2012vw,DeLuca:19}, while experimental measurements in gold nanoparticle arrays have demonstrated SHG efficiencies comparable to those in nonlinear crystals when normalized to the active volumes \cite{klein2006second}.

In this work, we present a method to study SHG originating from FE nonlinearities in plasmonic waveguides, within the context of the hydrodynamic theory.
Our method is based on writing the second-harmonic (SH) field along the waveguide as the energy flux provided by the nonlinear polarization field  at the waveguide cross-section, times the envelop due to the phase delay between the phases of the driving field and the waveguide modes at the SH frequency.
We then utilize our method to study SHG in distinct plasmonic waveguides based on semi-classical hydrodynamic nonlinearities, as well as a generalized quantum hydrodynamic theory with electron spill-out effects. 
We validate our method through  full-wave numerical simulations of SHG in a simple waveguide configuration.

\section{Theory}\label{II}
The hydrodynamic model has been extensively used to describe FE nonlinear dynamics in noble metals \cite{PhysRev.174.813,Khalid:2020bx, PhysRevB.79.235109,kauranen2012nonlinear,Scalora:2010kd,Ciraci:2012vw,krasavin2018free,DeLuca:19, PhysRevB.21.4389} and heavily doped semiconductors \cite{deluca2021}.
Within the hydrodynamic description, FE nonlinear dynamics, under the influence of an external electric, $\mathbf{E}(\mathbf{r},t)$, and magnetic $\mathbf{B}(\mathbf{r},t)$ fields can be described by the following equation \cite{Ciraci:2016il}:
\begin{eqnarray}
    m_{e}\left[\frac{\partial {\bf v}}{\partial t}+{\bf v}\cdot\nabla+\gamma \right]{\bf v}=e({\bf E}+{\bf v}\times{\bf B})-\nabla\frac{\delta G[n]}{\delta n}
    \label{me},
\end{eqnarray}
\noindent where $m_{e}$ is the electron mass, $\gamma$ is the electron collision rate, and $e$ is the elementary charge (absolute value).
The hydrodynamic variables $\mathbf{v}(\mathbf{r},t)$ and $n({\bf r},t)$ represent velocity and density of free electrons, respectively, and $G[n]$ contains the total internal energy of the electronic system \cite{Ciraci:2016il,Khalid:2020bx}.
The exact expression for $G[n]$ is unknown, however, it is possible to rely on approximated expression.
Its simplest form can be obtained in the Thomas-Fermi approximation, i.e., $G[n]=T_{\rm TF}[n]=(E_h a_0^2)c_{\rm TF}n^{5/3}$, where $E_h$ is the Hartree energy, $a_0$ is the Bohr radius and $ c_{\rm TF}=\frac{3}{10}(3\pi^2)^{2/3}$. This approach will be referred to as Thomas-Fermi hydrodynamic theory (TF-HT).

Eq.~\eqref{me} can be easily rewritten in terms of the polarization field $\mathbf{P}(\mathbf{r},t)$ considering that $\frac{\partial{\bf P}}{\partial t}=\mathbf{J}=-en{\bf v}$, where $\mathbf{J}(\mathbf{r}, t)$ is the current density.
Then, using a perturbative approach, it is possible to write $n({\bf r},t)=n_0+n_{\rm ind}({\bf r},t)$, where $n_0$ and $n_{\rm ind}=\frac{1}{e}\nabla \cdot{\bf P}$ are the equilibrium and the induced charge densities, respectively. For low enough excitation intensities $n_{\rm ind} \ll n_0 $, such that we can write:
\begin{eqnarray}
    -\beta^2\nabla\nabla\cdot{\bf P}+\frac{\partial^2{{\bf P}}}{\partial t^2}+\gamma\frac{{\partial{\bf P}}}{\partial t}=\varepsilon_0\omega_p^2{\bf E}+{\bf S}_{\rm NL}
    \label{eq:tildeP},
\end{eqnarray}
\noindent where $\beta^2=(E_h a_0^2)\frac{10}{9}\frac{c_{\rm TF}}{m_e}n_0^{2/3}$, and ${\bf S}_{\rm NL}$ is the second-order nonlinear source, including Coulomb, Lorentz, convective and nonlinear pressure terms \cite{Ciraci:2012vw}:
\begin{equation}
    \begin{split}
        \mathbf{S}_{\mathrm{NL}} =&\frac{e}{m_e} \mathbf{E}(\nabla \cdot \mathbf{P})-\frac{\mu_{0} e}{m_e}(\dot{\mathbf{P}} \times \mathbf{H})\\
        &+\frac{1}{n_{0} e}(\dot{\mathbf{P}} \nabla \cdot \dot{\mathbf{P}}+\dot{\mathbf{P}} \cdot \nabla \dot{\mathbf{P}})\\
        &+\frac{1}{3} \frac{\beta^{2}}{n_{0} e} \nabla(\nabla \cdot \mathbf{P})^{2}.
    \end{split}
    \label{eq:S_TD}
\end{equation}
Here, $\dot{\mathbf P}$ represents the time derivative of the polarization field.

In order to study SHG, let us expand the fields into two time-harmonic contributions, ${\bf{F}}({\bf{r}},t) = \sum\limits_j {{{\bf{F}}_j}({\bf{r}}){e^{ - i{\omega _j}t}}}$, with ${\bf{F}}={\bf{E}}$, ${\bf{H}}$, or ${\bf{P}}$ and $j= 1,2$. Eqs. (\ref{eq:tildeP}-\ref{eq:S_TD}) and Maxwell's equations can then be rewritten as a set of equations for each harmonic $\omega_j$:
\begin{subequations}
    \begin{gather}
        \nabla\times\nabla\times {\bf E}_j-k_j^2{\bf E}_j=\mu_0\omega_j^2{\bf P}_j,
        \label{eq:wave_SH}\\
        -\beta^2\nabla\nabla\cdot{\bf P}_j-\left(\omega_j^2+i\gamma\omega_j \right){\bf P}_j=\varepsilon_0\omega_p^2{\bf E}_j+{\bf S}_{j\rm, NL},
        \label{eq:P_SH}
        \end{gather}
        \label{eq:wave_P_SH}
\end{subequations}
where $k_j$ is the free-space wavenumber. Considering that $\nabla\cdot{\bf P}=-\varepsilon_0\nabla\cdot{\bf E}$, the polarization field can be expressed as a function of the electric field:
\begin{equation}
    {\bf P}_j=\varepsilon_0\chi_j\left[{\bf E}_j-\frac{\beta^2}{\omega_p^2}\nabla\nabla\cdot{\bf E}_j\right]+{\bf P}_{j,{\rm NL}},\label{eq:P2}
\end{equation}

\noindent where $\chi_j=\varepsilon_r(\omega_j)-1=-\frac{\omega_p^2}{\omega^2+i\gamma\omega_j}$ and ${\bf P}_{j,{\rm NL}}=\frac{\chi_j}{\omega_p^2}{\bf S}_{j,{\rm NL}}$.
Finally, from Eqs.~\eqref{eq:wave_P_SH}, we get the following system:
\begin{subequations}
    \begin{gather}
       \begin{split}
        \nabla^2 {\bf E}_1-\left[1-\beta^2\frac{k_1^2\chi(\omega_1)}{\omega_p^2}\right]\nabla\nabla\cdot{\bf E}_1 &\\
        +\varepsilon_r(\omega_1)k_1^2{\bf E}_1=0 &,
        \label{eq:wave_E1}
        \end{split}
    \\
        \begin{split}
            \nabla^2 {\bf E}_2-\left[1-\beta^2\frac{k_2^2\chi(\omega_2)}{\omega_p^2}\right]\nabla\nabla\cdot{\bf E}_2 &\\
            +\varepsilon_r(\omega_2)k_2^2{\bf E}_2=-\mu_0\omega_2^2{\bf P}_{2,{\rm NL}} &,
            \label{eq:wave_E2}   
        \end{split} 
    \end{gather}
\end{subequations}
\noindent where, for simplicity, under the undepleted pump approximation, we assumed ${\bf P}_{1,{\rm NL}}\simeq0$, while the second-order nonlinear source becomes:
\begin{equation}
    \begin{split}
        {\bf S}_{2,{\rm NL}}=&\frac{e}{m_e}{{\bf{E}}_1}\nabla  \cdot {{\bf{P}}_1} + i\frac{{{\omega _1}{\mu _0}e}}{m_e}{{\bf{P}}_1} \times {{\bf{H}}_1}\\ 
        & - \frac{{\omega _1^2}}{{{n_0}e}}\left[ {{\bf{P}}_1}(\nabla  \cdot {{\bf{P}}_1})+ {{\bf{P}}_1} \cdot \nabla {{\bf{P}}_1} \right] \\ 
        & + \frac{1}{3}\frac{{{\beta ^2}}}{{{n_0}e}}\nabla {\left( {\nabla  \cdot {{\bf{P}}_1}} \right)^2}.
        \label{eq:S2NL}
    \end{split}
\end{equation}

% Eqs. \eqref{eq:wave_E1} and \eqref{eq:wave_E2} can be solved assuming hard-wall boundary conditions, that is the electrons are not allowed to escape the metal boundary, i.e., $\mathbf{P} \cdot \hat{\mathbf{n}}=0$ \cite{Ciraci:2016il}, at the metal surface.

Eqs. \eqref{eq:wave_E1} and \eqref{eq:wave_E2} can be solved assuming the continuity of the normal component of the polarization vector, i.e., $P_{n}^{-}=P_{n}^{+}$.
This assumption is often combined with a constant equilibrium density $n_0$ in the metal, while being zero outside (hard-wall boundary conditions) \cite{Raza:2013bd,Huang:13,Toscano:2013hr,krasavin2018free,Scalora:2010kd,Ciraci:2012vw}.

We are interested in waveguide solutions at this point. In order to derive the fundamental field (FF) from Eq.~\eqref{eq:wave_E1}, let us assume, without loss of generality, that the modes propagate along the $z$-direction. The solution is then of the form ${\bf E}_1({\bf r})=A_{1}\tilde{\bf E}_1(x,y)e^{i\kappa_1 z}$,
where $\kappa_1$ is the complex mode propagation constant, $A_{1}$ is the mode amplitude, and $\tilde{\bf E}_1(x,y)$ is the mode profile of the FF at the waveguide cross-section. 
By writing $\nabla=\nabla_\perp+i\kappa_1\hat{\bf z}$, Eq.~\eqref{eq:wave_E1} can be solved either analytically, in a few simple cases \cite{Raza:2013bd}, or numerically, for an arbitrary waveguide cross-section \cite{Toscano:2013hr,Huang:13,Zheng.2019}, using an eigenmode solver to calculate mode profile and propagation constant. The propagation constant of the mode is defined as: 
\begin{equation}
    \kappa=\beta+i\frac{\alpha}{2}
    \label{wn}
\end{equation}
\noindent with $\beta$ and $\alpha$ being the propagation and attenuation constant of the mode, respectively. 
In our implementation we have used \textit{Comsol Multiphysics} \cite{comsol} with a customized weak form.
The found mode can then be normalized assuming the input-power at the $z=0$ waveguide cross-section to be 1 W:

\begin{equation}
    %A_{1}^2
    \frac{1}{2}\int_\Omega{\rm Re}\left[\tilde{\bf E}_1\times \tilde{\bf H}_1^*\right]\cdot\hat{\bf z}dS=1 {\rm W},
    \label{eq:input-power}
\end{equation}

where $\Omega$ is the cross-sectional plane. Therefore, within these assumptions, the second-order nonlinear source in Eq.~\eqref{eq:wave_E2} can be rewritten as:

\begin{eqnarray}
    \begin{split}
        {\bf P}_{2,{\rm NL}}({\bf r})= & A_1^2\frac{\chi(\omega_2)}{\omega_p^2}e^{2i\kappa_1 z}\left\{\frac{e}{m_e}{\tilde{\bf{E}}_{1}}(\nabla_\perp+i\kappa_1\hat{\bf z})  \cdot {\tilde{\bf{P}}_{1}}\right. \\
        &+ i\frac{\omega_{1}\mu _0 e}{m_e}\tilde{\bf{P}}_{1} \times\tilde{\bf{H}}_{1} \\
        &-\frac{\omega _1^2}{n_0e}\left[ {\tilde{\bf{P}}_{1}}\left[(\nabla_\perp+i\kappa_1\hat{\bf z})  \cdot {\tilde{\bf{P}}_{1}}\right] \right. \\ 
        & + \left.\tilde{\bf P}_{1} \cdot (\nabla_\perp+i\kappa_1\hat{\bf z}) \tilde{\bf P}_1 \right] \\
        &\left.+\frac{1}{3}\frac{\beta ^2}{n_0e}(\nabla_\perp+i\kappa_1\hat{\bf z}) \left[(\nabla_\perp+i\kappa_1\hat{\bf z})  \cdot \tilde{\bf P}_1 \right]^2\right\},
        \label{eq:P2NL}
    \end{split}
\end{eqnarray}

\noindent where the mode is normalized in such a way that $A_1^2$ is the pump input power.

For the SHG, let us now consider Eq. \eqref{eq:wave_E2}. In nonlinear optics, the divergence term is generally neglected and a solution of Eq. \eqref{eq:wave_E2} can be easily obtained in the slowly varying envelope approximation, through the definition of overlap integrals evaluated in the waveguide cross-section \cite{Ruan:09,davoyan2010backward,Zhang:13,Zhang:13a,Wu:14,Sun:15,Huang:16,shi2019efficient}.
In the case of metal nonlinearities, and in particular of hydrodynamic nonlinearities, neglecting the divergence will strongly affect the results, since the larger nonlinear contributions arise at the metal surface, where the divergence is non-zero.
On the other hand, fully solving Eq. \eqref{eq:wave_E2} in a three-dimensional numerical set-up is challenging, due to the large scale mismatch between the surface effects and the overall mode propagation. 

In what follows, we describe a procedure that allows to calculate SHG along the waveguide by only solving a numerical problem on a two-dimensional cross-section of the waveguide.

The general solution of the partial differential equation \eqref{eq:wave_E2} is given by the sum of the solution of the homogeneous equation (i.e., assuming ${\bf P}_{2,{\rm NL}}({\bf r})=0$) and a particular solution of the inhomogeneous equation, i.e., ${\bf E}_2({\bf r})={\bf E}_h({\bf r})+{\bf E}_p({\bf r})$. 
${\bf E}_h({\bf r})=\sum_m a_m\tilde{\bf E}_m(x,y)e^{i\kappa_mz}$ with $\tilde{\bf E}_m$ being the modes supported by the waveguide at $\omega_2$, and $a_m$ are amplitude coefficients to be determined.
The modes $\tilde{\bf E}_m$ can be easily found through an eigenmode solver.
As usual, we assume that the modes are normalized to carry the same input power, i.e.,

\begin{equation}
    \frac{1}{2}\int_\Omega \tilde{\bf E}_m\times\tilde{\bf H}^*_m\cdot \hat{\bf z}dS=1{\rm W},
    \label{WM}
\end{equation}

Because the system is not lossless, the modes need to satisfy the following orthogonality relation \cite{mcisaac1991mode,mahmoud1991electromagnetic}:

\begin{equation}
    \int_\Omega\left(\tilde{\bf E}_m\times \tilde{\bf H}_n\right)\cdot\hat{\bf z}dS=N_m\delta_{nm},
    \label{NM}
\end{equation}

\noindent with

\begin{equation}
    N_m=\int_\Omega\left(\tilde{\bf E}_m\times \tilde{\bf H}_m\right)\cdot\hat{\bf z}dS.
    \label{eq:ortho}
\end{equation}

The particular solution can be sought of the form ${\bf E}_p({\bf r})=\tilde{\bf E}_p(x,y)e^{i2\kappa_1 z}$ where $\kappa_1$ is the known FF's propagation constant.
Eq. \eqref{eq:wave_E2} then can be solved in the waveguide cross-section by transforming the nabla operator as $\nabla=\nabla_\perp+2i\kappa_1\hat{\bf z}$.
Once ${\bf E}_p({\bf r})$ is known we can determine the coefficients $a_m$ by imposing the total power flow to be zero at the waveguide input, $z=0$:

\begin{equation}
    W_{\rm SHG}(z=0)=\frac{1}{2}\int_\Omega{\rm Re}\left[{\bf E}_2\times {\bf H}_2^* \right]\cdot\hat{\bf z}dS=0.
    \label{eq:WSHG_zero}
\end{equation}

In order to do so, it is useful to project the field ${\bf E}_p$ on the waveguide modes at $z=0$, i.e., find the coefficients $b_m$ such that ${\bf E}_p(z=0)=\sum_m b_m\tilde{\bf E}_m$.

These coefficients can be found as \cite{mcisaac1991mode,mahmoud1991electromagnetic}:

\begin{equation}
    b_m=\frac{1}{2N_m}\int_\Omega\left(\tilde{\bf E}_p\times \tilde{\bf H}_m+\tilde{\bf E}_m\times \tilde{\bf H}_p\right)\cdot\hat{\bf z}dS.
    \label{eq:bm}
\end{equation}

The condition of Eq. \eqref{eq:WSHG_zero} then becomes:

\begin{equation}
    \begin{split}
         \mathlarger{\mathlarger{\sum}}_{m,n}\bigg[\left(a_{m} a_{n}^{*}+a_{m} b_{n}^{*}+b_{m} a_{n}^{*}+b_{m} b_{n}^{*}\right)\bigg.& \\ \left.\times\int_\Omega\left(\tilde{\bf E}_m\times \tilde{\bf H}_n^* \right)\cdot\hat{\bf z}dS\right]=0&,
    \end{split}
    \label{eq:WSHG_zero2}
\end{equation}

If the number of modes and losses are small such that $\sum_{m\ne n}\int_\Omega\left(\tilde{\bf E}_m\times \tilde{\bf H}_n^* \right)\cdot\hat{\bf z}dS\ll\sum_{m}\int_\Omega\left(\tilde{\bf E}_m\times \tilde{\bf H}_m^* \right)\cdot\hat{\bf z}dS$, Eq.~\eqref{eq:WSHG_zero2} can be simplified as:

\begin{equation}
    \begin{split}
        \mathlarger{\mathlarger{\sum}}_{m}\bigg[\left(\left|a_{m}\right|^{2}+a_{m} b_{m}^{*}+b_{m} a_{m}^{*}+\left|b_{m}\right|^{2}\right)\bigg.& \\ \left.\times\int_\Omega\left(\tilde{\bf E}_m\times \tilde{\bf H}_m^* \right)\cdot\hat{\bf z}dS\right]\simeq0&,
    \end{split}
    \label{eq:WSHG_zero3}
\end{equation}

Since the quantity in the integral is nonzero it must be:

\begin{equation}
    \sum_{m}\left(|a_m|^2+a_mb_m^*+b_ma_m^*+|b_m|^2\right)=0.
    \label{eq:WSHG_zero4}
\end{equation}

Eq.~\eqref{eq:WSHG_zero4} can be satisfied by choosing $a_m=-b_m$.
The SH field then can be written as:

\begin{equation}
    {\bf E}_2({\bf r})=\sum_m b_m\tilde{\bf E}_m(x,y)\left(e^{i2\kappa_1z}-e^{i\kappa_mz}\right),
\end{equation}

and the SHG power as a function of the propagation distance $z$ is given by:

\begin{eqnarray}
W_{\rm SHG}(z)= \sum_m|b_m|^2|e^{i2\kappa_1z}-e^{i\kappa_mz}|^2.\label{eq:W_SHG}
\end{eqnarray}

Equation~\eqref{eq:W_SHG} constitutes the main result of this section. The SHG power along the waveguide can be obtained through the mode propagation constants, $\kappa_1$ and $\kappa_m$, at the FF and SH wavelengths, respectively.
Note that if only one mode is supported by the waveguide at $\omega_2$, i.e. $b_1=b$, then $|b|^2=\frac{1}{2}\int_\Omega{\rm Re}\left[\tilde{\bf E}_p\times \tilde{\bf H}_p^* \right]\cdot\hat{\bf z}dS$.
In the following, we will refer to this method as the \textit{particular solution method} (PSM).

\section{Results}\label{III}
In this section, we present some application examples of SHG in waveguides with hydrodynamic nonlinearities.
In order to validate our method, we first consider a simple metal-insulator-metal (MIM) waveguide.
Because of the translation symmetries of the system, in fact, it is possible to easily perform full-wave calculations without having to rely on a three-dimensional implementation of the hydrodynamic equations \cite{Vidal-Codina.2021}.
Subsequently, we apply the PSM to a typical waveguide design without any translation symmetry in the transverse plane.
Finally, we demonstrate the validity of the PSM for a system in which electron spill-out effects are taken into account through a more sophisticated model. 

\subsection{Second-harmonic generation in metal-insulator-metal waveguides} \label{A}

\begin{figure}[!tb]
  \begin{center}
    \includegraphics[width=3.3in]{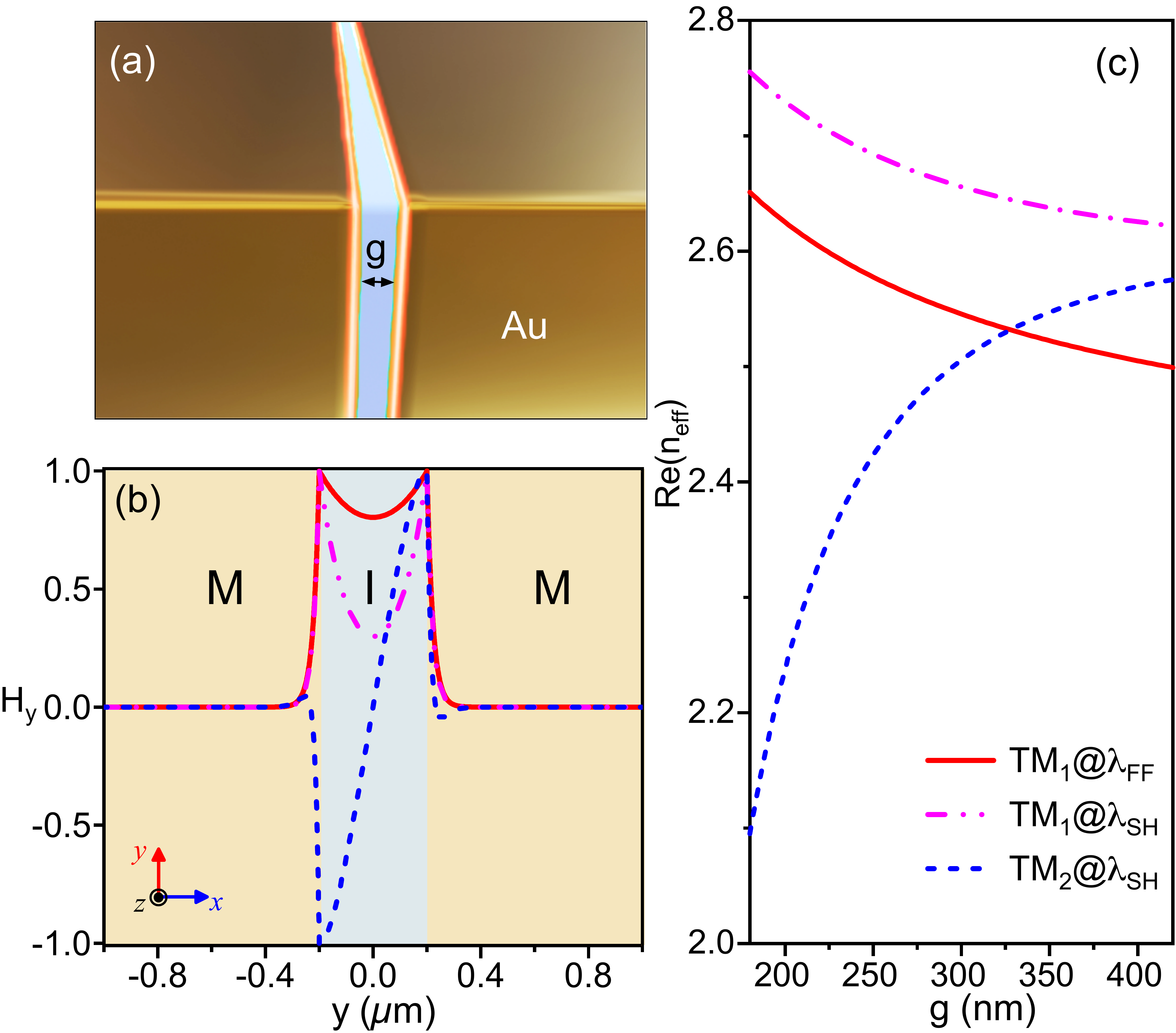}
  \caption{\small The MIM waveguide: (a) schematic of the geometry, (b) magnetic field profiles, and (c) real part of the effective refractive indices as a function of the gap size of the supported modes.}
  \label{fig1}
\end{center}
\end{figure}

Different types of metal-dielectric waveguides have been presented theoretically and demonstrated experimentally (see, e.g., Refs. \cite{maier2007plasmonics,economou1969surface,bruke,prade}). 
Here, we study a symmetric configuration, i.e., a thin dielectric layer of size \textit{g} sandwiched between two gold surfaces (with the metal extending indefinitely on both sides of the dielectric), as shown in
\crefformat{figure}{Fig.~#2#1{(a)}#3}\cref{fig1}.
We consider the following parameters for gold: $n_0=5.7\times10^{22}~{\rm cm}^{-3}$, $\gamma=1.07\times10^{14}~{\rm s}^{-1}$, and $\beta=1.27\times10^{6}~{\rm ms^{-1}}$ \cite{Ciraci:2012vw}, while the dielectric layer has a relative permittivity $\varepsilon_{d}=5.56$. The wavelengths considered for parametric interaction are $\lambda_{\rm FF}=1550~{\rm nm}$ and $\lambda_{\rm SH}=775~{\rm nm}$ at the FF and SH, respectively. 
The MIM waveguide supports symmetric gap-plasmon modes at both FF and SH wavelengths, denoted as TM$_1@\lambda_{\rm FF/SH}$, and an anti-symmetric SPP at SH, indicated as TM$_2@\lambda_{\rm SH}$ (see \crefformat{figure}{Fig.~#2#1{}#3}\cref{fig1}). We render the magnetic field profiles and real part of the effective indices of the modes in \crefformat{figure}{Fig.~#2#1{(b)}#3}\cref{fig1} and \crefformat{figure}{Fig.~#2#1{(c)}#3}\cref{fig1}, respectively.
As shown in the latter figure, their dispersive behavior holds for a wide range of gap sizes.

An efficient energy transfer from the mode at the FF to that at the SH can be obtained if a gap size is chosen that guarantees a phase-matching (PM) condition \cite{Zhang:13,Zhang:13a,Wu:14,Sun:15,Huang:16,Davoyan:09}.
In our case, as it can be seen in \crefformat{figure}{Fig.~#2#1{(c)}#3}\cref{fig1}, the PM occurs between the symmetric mode TM$_1@\lambda_{\rm FF}$ at FF and the higher-order anti-symmetric modes TM$_2@\lambda_{\rm SH}$ at the SH wavelength for a gap size of $g\approx327~{\rm nm}$. 
\begin{figure}[!tb]
  \begin{center}
    \includegraphics[width=2.8in]{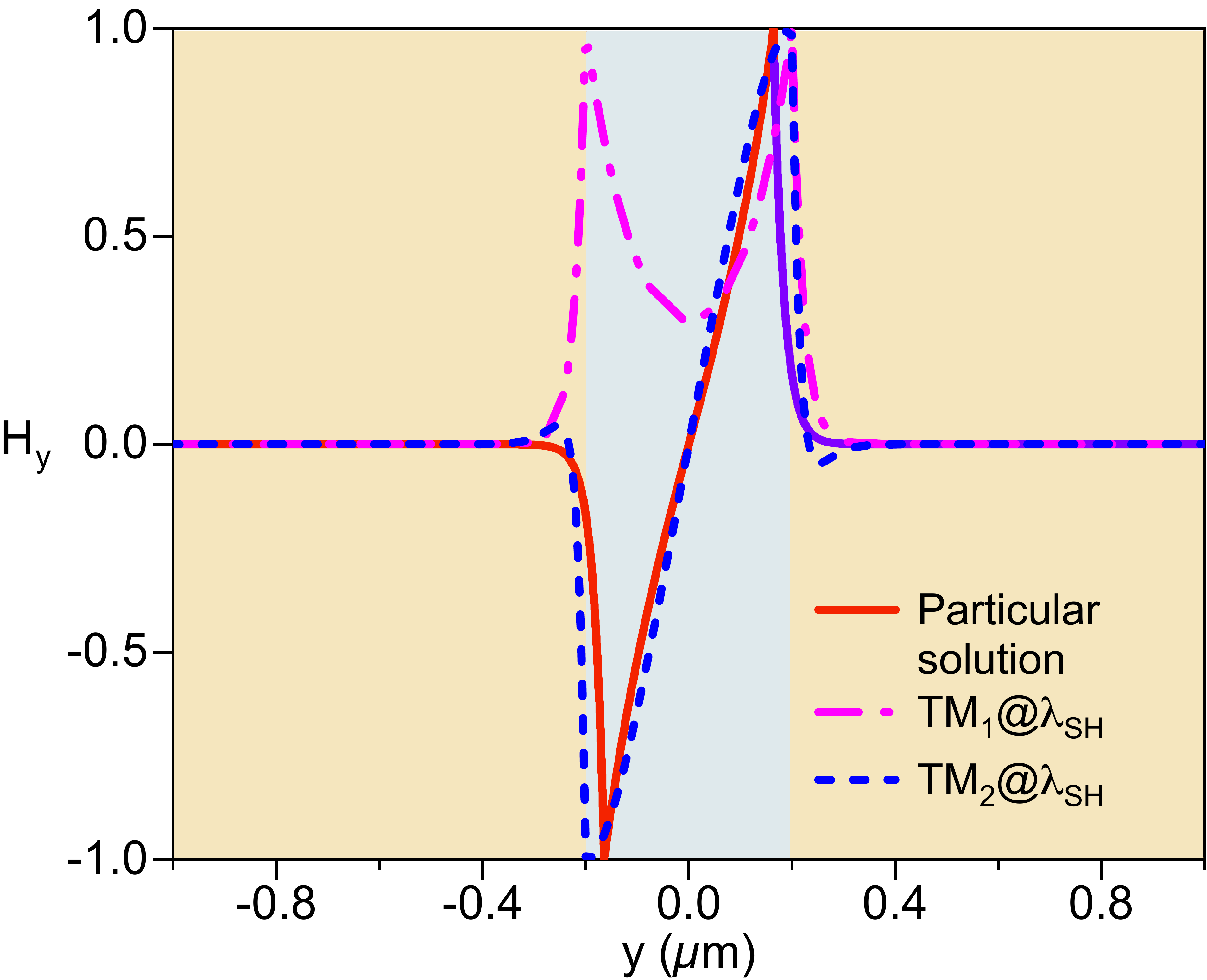}
  \end{center}
\caption{Magnetic field transverse component, $H_y$, of the particular solution and modes at the SH wavelength for the MIM waveguide.}
  \label{fig2}
\end{figure}
For the validation of our method we consider two situations: i) the just mentioned phase-matched case, and ii) a non-phase-matched, with $g=270$~nm. 
We assume that the whole FF energy is in the TM$_1@\lambda_{\rm FF}$ mode, while the SHG can couple to both TM$_1@\lambda_{\rm SH}$ and TM$_2@\lambda_{\rm SH}$.
In \crefformat{figure}{Fig.~#2#1{}#3}\cref{fig2} we show the magnetic field profile of the particular solution (PS) obtained by considering the nonlinear polarization in Eq.~\eqref{eq:P2NL}, as well as the modes available at the SH.
It is easy to guess from the plot that most of the SHG energy will be coupled to TM$_2@\lambda_{\rm SH}$, due to the modes' symmetries.
Indeed, this is confirmed by the evaluation of the coefficients $|b_m|^2$ associated to the modes, which differ by several orders in magnitude (see Table \ref{table:MI}).
\begin{table}[ht]
\caption{Coefficients $|b_m|^2$ and energy flux $W_{\rm p}$ of the particular solution for the MIM waveguide.}
\renewcommand{\arraystretch}{1.5}
\begin{tabular}{c|cc|c}
\cline{1-4}
$g$ (nm)             & $|b_{\rm TM_1}|^2$ & $|b_{\rm TM_2}|^2$ & $W_{\rm p}$ (W)  \\ \cline{1-4}
327     & $7.6\times10^{-22}$     & $0.25$               & $0.25$    \\
270 & $2.9\times10^{-24}$     & $2.2\times10^{-3}$    & $2.2\times10^{-3}$ \\ \cline{1-4}
\end{tabular}
\label{table:MI}
\end{table}

\begin{figure}[!htb]
  \begin{center}
    \includegraphics[width=3.3in]{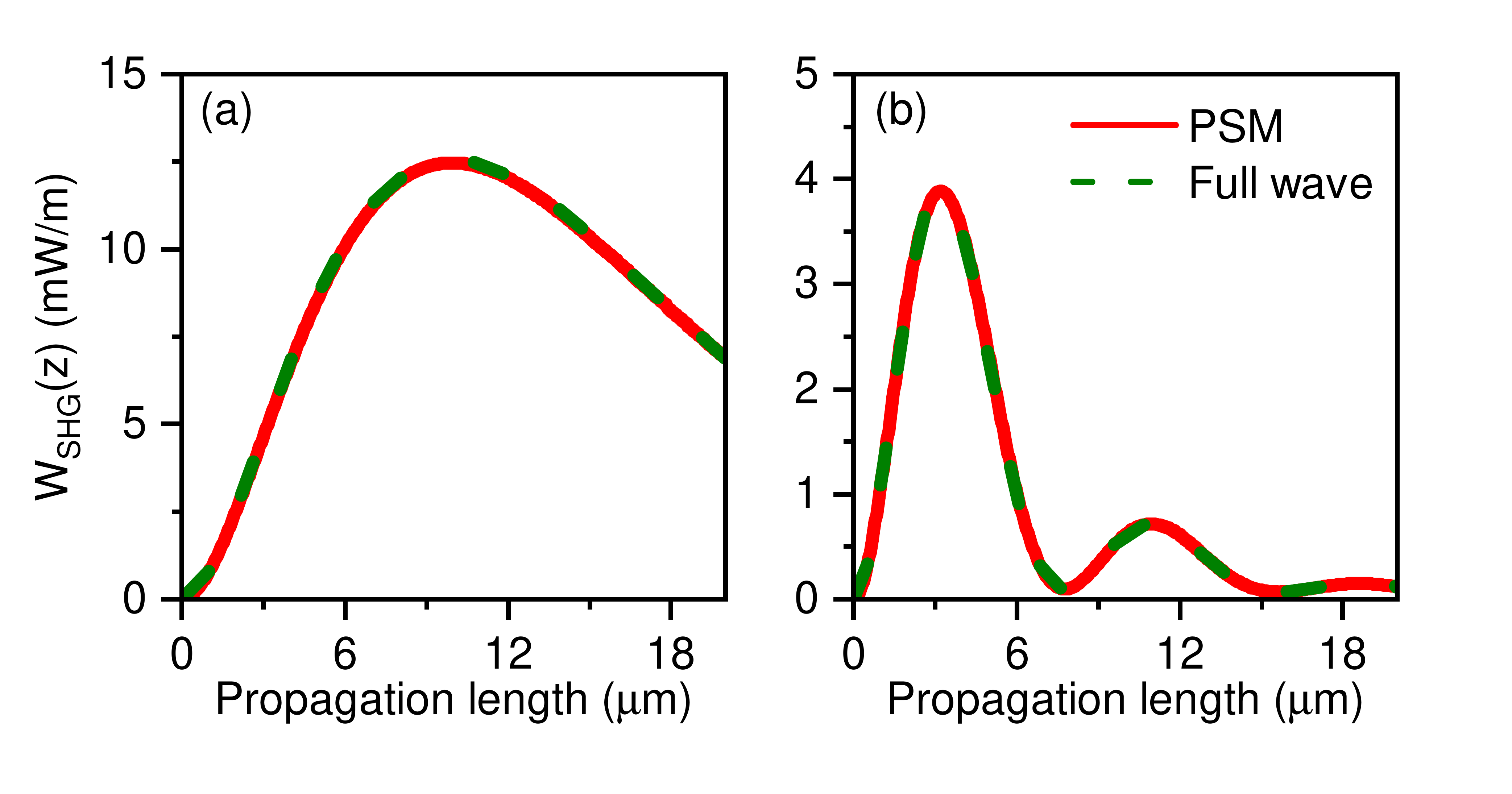}
  \end{center}
\caption{\small SHG intensity as function of the of the propagation distance: (a) the phase-matched and (b) non phase-matched case for the MIM waveguide. Wavenumbers of the interacting TM$_1@\lambda_{\rm FF}$ and TM$_2@\lambda_{\rm SH}$ modes are: $\kappa_{\rm FF}=1.02\times10^7+3.64\times10^{4}i$, $\kappa_{\rm SH}=2.05\times10^7 + 1.34\times10^5i$ in (a) and $\kappa_{\rm FF}=1.03\times10^7 + 4.19\times10^{4}i$, $\kappa_{\rm SH}=1.99\times10^7 + 1.49 \times10^5i$ in (b).}
  \label{fig3}
\end{figure}

By using Eq.~\eqref{eq:W_SHG} we can calculate the SHG power along the waveguide, reported in \crefformat{figure}{Fig.~#2#1{}#3}\cref{fig3} for the two studied cases, considering an input power of 1 MW/m.
As expected, in the phase-matched case we observe the SH signal building up until the losses in both the FF and the SH modes start affecting the conversion process. 
The SHG peak is obtained at approximately $10~{\rm \mu m}$.
Conversely, in the non-phase-matched case, the SHG is limited first by the short coherence length, and then by the metal losses. 
However, in both cases we obtained perfect agreement with full-wave calculations \cite{Ciraci:2012vw,Scalora:2010kd,deluca2021}, performed by solving directly Eqs. \eqref{eq:wave_P_SH} in the $x$-$z$ plane (see \crefformat{figure}{Fig.~#2#1{}#3}\cref{fig3}).
These results shall lay a foundation for the applicability of the PSM to characterize the SHG in a variety of waveguides with hydrodynamic nonlinearities, as will be shown in the following subsections.
\subsection{Non-planar waveguide with hydrodynamic nonlinearities}
\begin{figure}[!tb]
  \begin{center}
    \includegraphics[width=3.3in]{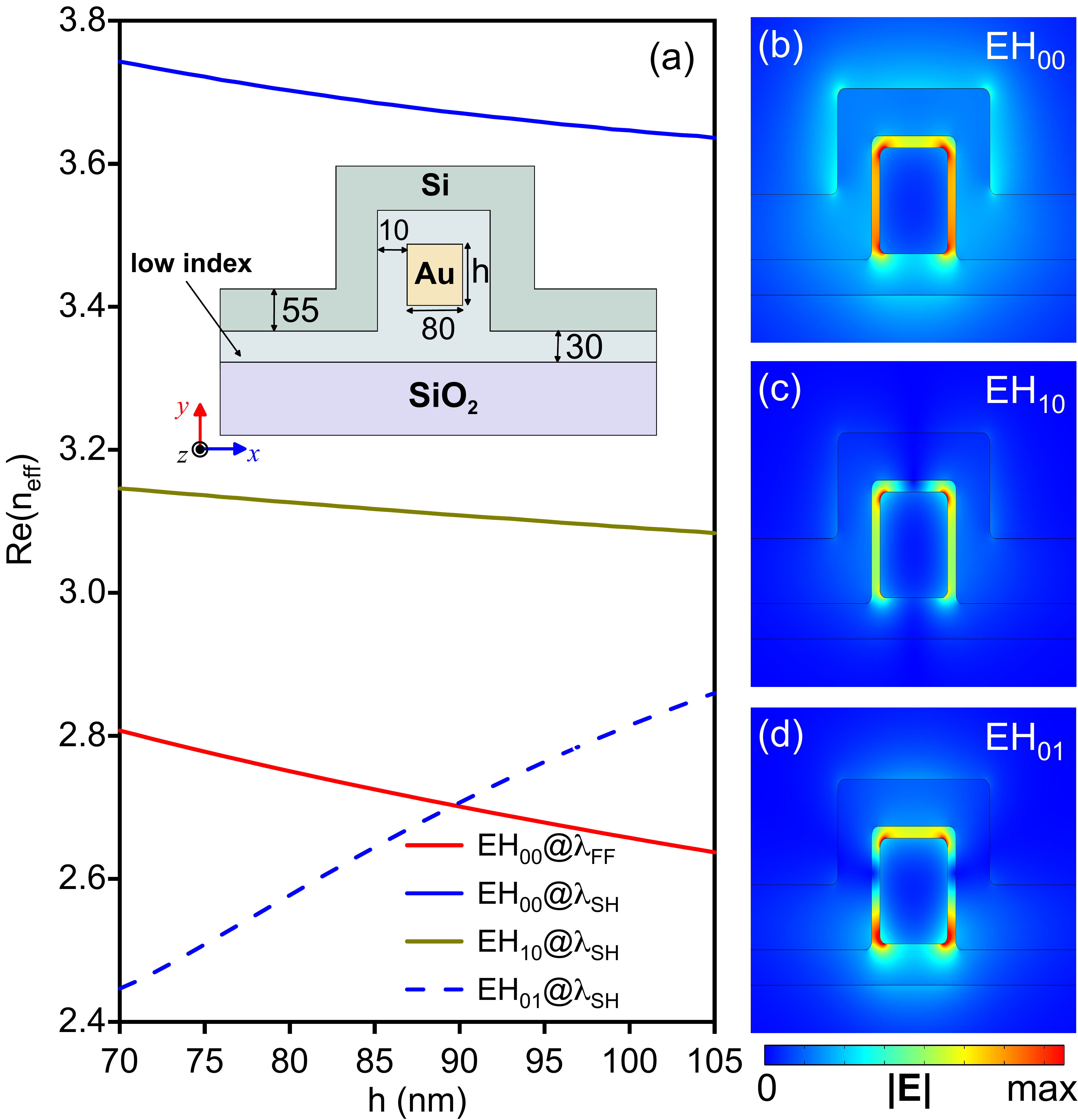}
  \end{center}
\caption{\small Non-planar waveguide: (a) Tuning of the effective refractive indices as a function of the height of the metallic core, h, and (b-d) electric field profiles of modes supported by the structure at distinct wavelengths; the inset in (a) shows a schematic of the non-planar waveguide; dimensions are in nanometers. The permittivities used are: $\varepsilon_{\rm Si}=12.25$, $\varepsilon_{\rm SiO_2}=2.0$ and $\varepsilon_l=3.422$ for the low-index dielectric.}
  \label{fig4}
\end{figure}

Non-planar waveguides, characterized by an index profile $n$ that is a function of both transverse coordinates,are the most used in device applications
There are many examples of this kind of structures, differentiated by the distinctive features of their index profiles \cite{ZIA200620,yang2012subwavelength,Han_2012,fang2015nanoplasmonic}.
Here, we consider a non-planar waveguide whose cross-section is shown in the inset of \crefformat{figure}{Fig.~#2#1{(a)}#3}\cref{fig4}, together with its dispersion characteristics.
The structure consists of a ridge made of high-index dielectric material (Si) grown over a rectangular nanowire metallic core (which will act as a nonlinear medium) surrounded by a low-index dielectric material placed on top of a SiO$_2$ substrate.
The index contrast of the waveguide's constituents enforces the electromagnetic energy to be confined in the core-region of the ridge, which can be exploited to enhance nonlinearities present in that region while reducing losses associated to a typical plasmonic waveguide.

The waveguide is designed to support the FF mode at $\lambda_1=1300$~nm, while generating at $\lambda_2=650$~nm.
We present the modal structure of the waveguide in \crefformat{figure}{Fig.~#2#1{}#3}\cref{fig4}.
The variation of the mode effective indices as a function of the height $h$ of the metallic core is reported in \crefformat{figure}{Fig.~#2#1{(a)}#3}\cref{fig4}, while the norm of the electric field of the supported modes is shown in \crefformat{figure}{Fig.~#2#1{(b-d)}#3}\cref{fig4}. We observe that a lower-order hybrid mode of the non-planar waveguide appears at both the FF and SH wavelength (see the trends EH$_{00}@\lambda_{\rm FF/SH}$ in \crefformat{figure}{Fig.~#2#1{}#3}\cref{fig4}).
Whereas, the modal dispersion of the guided modes dictates that the higher-order hybrid modes indicated as EH$_{10/01}@\lambda_{\rm SH}$ are excited only at the SH wavelength.
The PM condition occurs between the EH$_{00}@\lambda_{\rm FF}$ and EH$_{01}@\lambda_{\rm SH}$ for h=89.5 nm, for fixed geometrical parameters (see the inset of \crefformat{figure}{Fig.~#2#1{(a)}#3}\cref{fig4}).  
   
\begin{figure}[!tb]
  \begin{center}
    \includegraphics[width=3.3in]{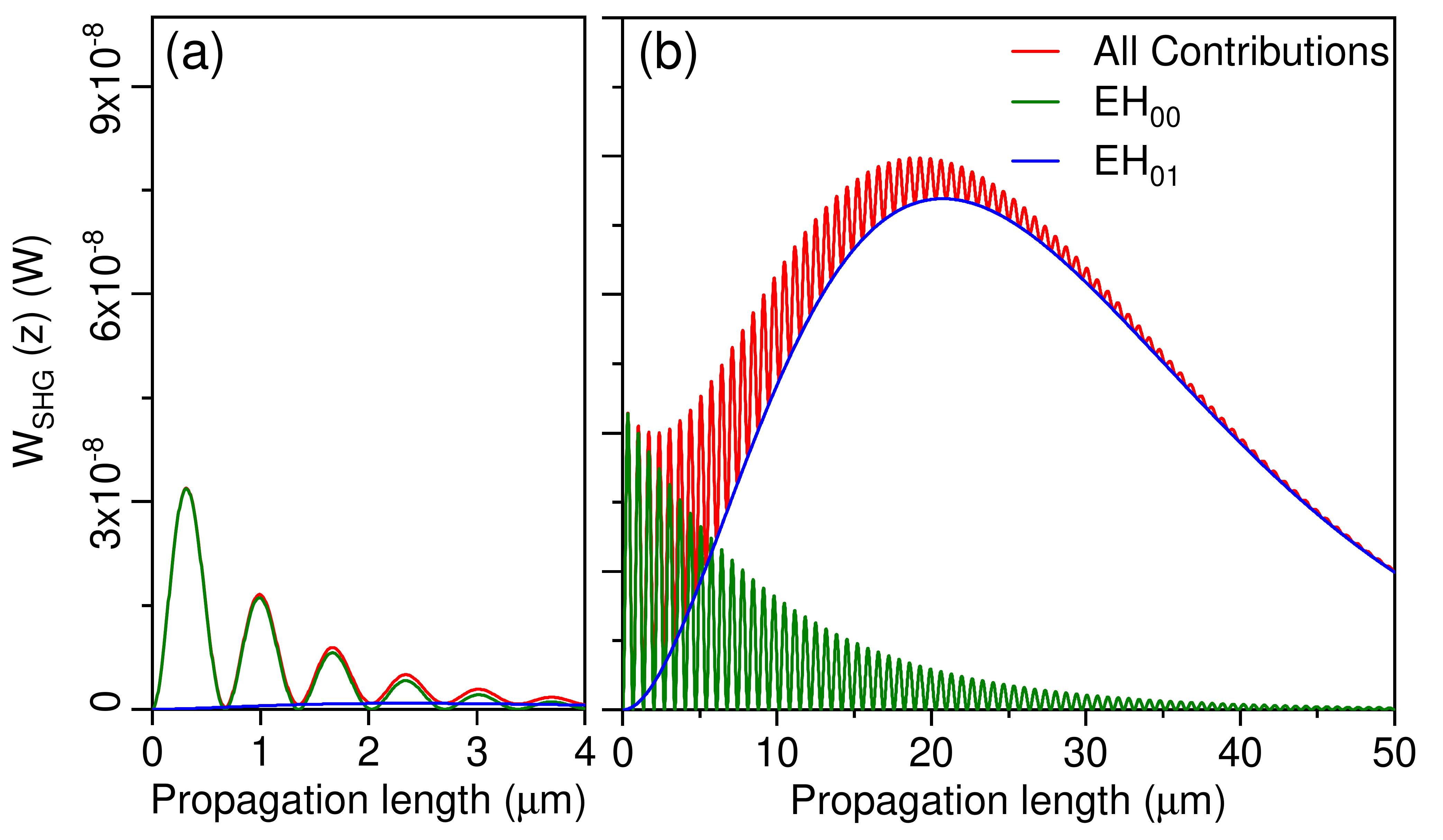}
  \end{center}
\caption{\small Evolution of SHG intensity along the non-planar waveguide, (red line) total and individual SH modes’ contributions (green line) EH$_{00}$ and (blue line) EH$_{01}$: (a) original system and (b) the case of reduced losses. The wavenumbers of the mode involved are: $\kappa_{\rm FF}=1.30\times10^7+2.38\times10^5i$, $\kappa_{\rm EH_{00}}=3.53\times10^7+5.34\times10^5i$ and $\kappa_{\rm EH_{01}}=2.60\times10^7+3.84\times10^5i$.}
  \label{fig5}
\end{figure}

Let us consider a pump input power of 1 W and start quantifying the contribution of each of the mode at the SH interaction wavelength to the SHG.
Based on the calculated $|b_{m}|^2$ of each of the mode at SH wavelength, we conclude that both the modes EH$_{00}@\lambda_{\rm SH}$ and EH$_{01}@\lambda_{\rm SH}$ can contribute to the SHG (see Table \ref{table:nonplan}).

\begin{table}[ht]
\caption{Coefficients $|b_m|^2$ and energy flux $W_{\rm p}$ of the particular solution for the non-planar waveguide.}
\renewcommand{\arraystretch}{1.5}
\begin{tabular}{c|ccc|c}
\hline
$\gamma$  (s$^{-1}$)          & $|b_{\rm EH_{00}}|^2$ & $|b_{\rm EH_{10}}|^2$ & $|b_{\rm EH_{01}}|^2$ &  $W_{\rm p}$ (W)     \\ \hline
$1.07\times10^{14}$ & $1.1\times10^{-8}$        & $1.5\times10^{-18}$        & $9.1\times10^{-8}$        & $1.0\times10^{-7}$ \\
$1.07\times10^{13}$  & $1.1\times10^{-8}$        & $1.5\times10^{-18}$        & $3.4\times10^{-7}$        & $3.5\times10^{-7}$ \\
\hline
\end{tabular}
\label{table:nonplan}
\end{table}

The single mode contributions and the total SHG power as a function of the propagation distance are reported in \crefformat{figure}{Fig.~#2#1{(a)}#3}\cref{fig5}.
Interestingly, the phase-matched mode (blue line) contributes almost negligibly to the overall SHG energy, which couple mostly into the non-phase-matched mode (green line).
This counterintuitive result is due to the interplay between the waveguide losses and the SHG build-up speed.
To understand this mechanism, let us artificially reduce the metal losses in the waveguide by one order in magnitude.
SHG along the waveguide length for such case is shown in \crefformat{figure}{Fig.~#2#1{(b)}#3}\cref{fig5}.
We observe that, although at small propagation distances, the non phase-matched EH$_{00}@\lambda_{\rm SH}$ carries more SHG energy than the phase-matched mode, it diminishes quickly, whereas the contribution from the phase-matched mode slowly builds up, peaking at a distance of around 25 $\mu$m.
We partially retrieve then the results for the ideal case without losses in which the SHG in the phase-matched mode increases until saturation of the pump. 
This example shows that, in general, the optimal device length is not determined by the coherence length of the phase-matched mode but it requires evaluating the contributions of all relevant modes. 
This is particularly relevant with hydrodynamic nonlinearities since most of the surface contributions drive strong evanescent fields that can easily couple to non-phase-matched modes.
\subsection{Electron spill-out}
In this section, we demonstrate the generality of the PSM by incorporating electron spill-out at the metal surfaces.
In writing Eq.~\eqref{eq:tildeP}, we assumed a specific approximation for the energy functional $G[n]=T_{\rm TF}[n]$, \textit{i.e.}, the Thomas-Fermi approximation with the hard-wall boundary conditions (i.e., no electron spill-out).
In the following, we express the functional in a more general form: $G[n]=T_{\rm TF}[n]+T_{\rm vW}[n, \nabla n]+E_{\rm XC}[n]$, where $T_{\rm vW}$ is the von Weizs\"acker correction to the TF kinetic energy and $E_{\rm XC }$ is the exchange-correlation energy functional.
The $\nabla n$-dependent correction in the kinetic energy functional allows to take into account the electron spill-out (spatial variation of charge density) at the metal interface.
This approach is generally known as quantum hydrodynamic theory (QHT). 

Eq. \eqref{eq:wave_E2} can be then be generalized to:
\begin{align}
    \begin{split}
        -\nabla\times\nabla\times {\bf E}_j&-\frac{\chi(\omega_j)k_j^2}{e}\nabla\left(\frac{\delta G[n]}{\delta n}\right)_j\\
        &+\varepsilon_r(\omega_j)k_0^2{\bf E}_j=-\mu_0\omega_j^2{\bf P}_{j,{\rm NL}}
        \label{eq:qht_Ej}
    \end{split}
\end{align}
where $j=1,2$ and ${\bf P}_{1,{\rm NL}}=0$ (undepleted pump approximation). The nonlinear polarization ${\bf P}_{2,{\rm NL}}$ must be enriched with nonlinear terms associated to the space dependent density as well as to the more complex expression of $G[n]$.
Detailed expressions for the linear functionals and ${\bf P}_{2,{\rm NL}}$ can be found in Refs. \cite{Ciraci:2016il, Khalid:2020bx}.

\begin{figure}%[!htb]
  \begin{center}
    \includegraphics[width=3.3in]{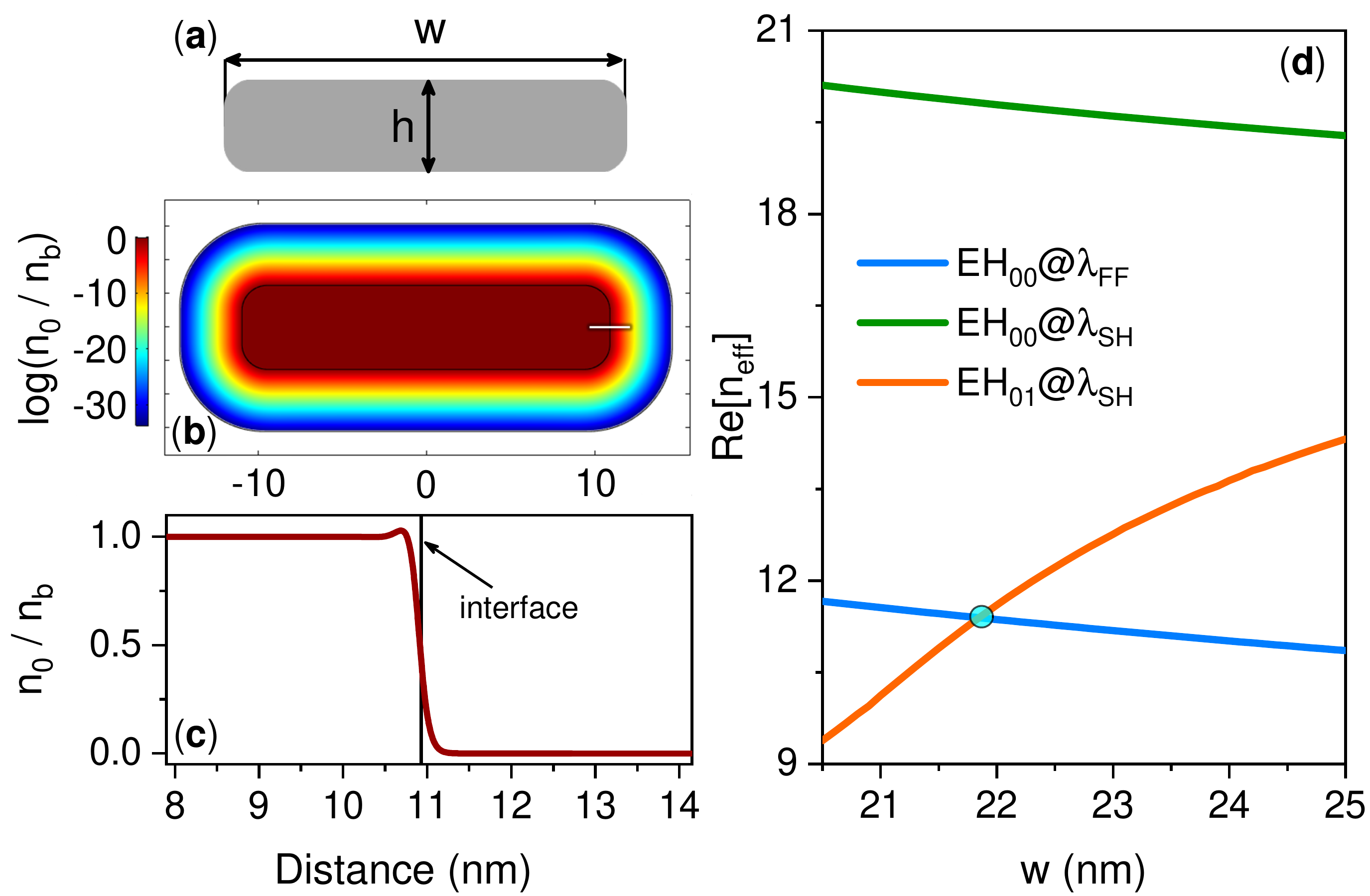}
  \end{center}
\caption{\small (a) Schematic of the strip waveguide embedded in a dielectric medium with $\varepsilon_d=5.56$. The sharp corners are rounded off with a radius of curvature of 1.5 nm. (b) The equilibrium charge density $n_0(\mathbf{r})$ normalized by the charge density in the bulk, $n_b$ and (c) the density profile near the metal-dielectric interface along the white line shown in (b). (d) Real part of $n_{\rm eff}$ as a function of the guide width $w$, considering $h=5$~nm.}
  \label{Fig:SO_Fig1}
\end{figure}
\begin{figure}%[!htb]
  \begin{center}
    \includegraphics[width=3.3in]{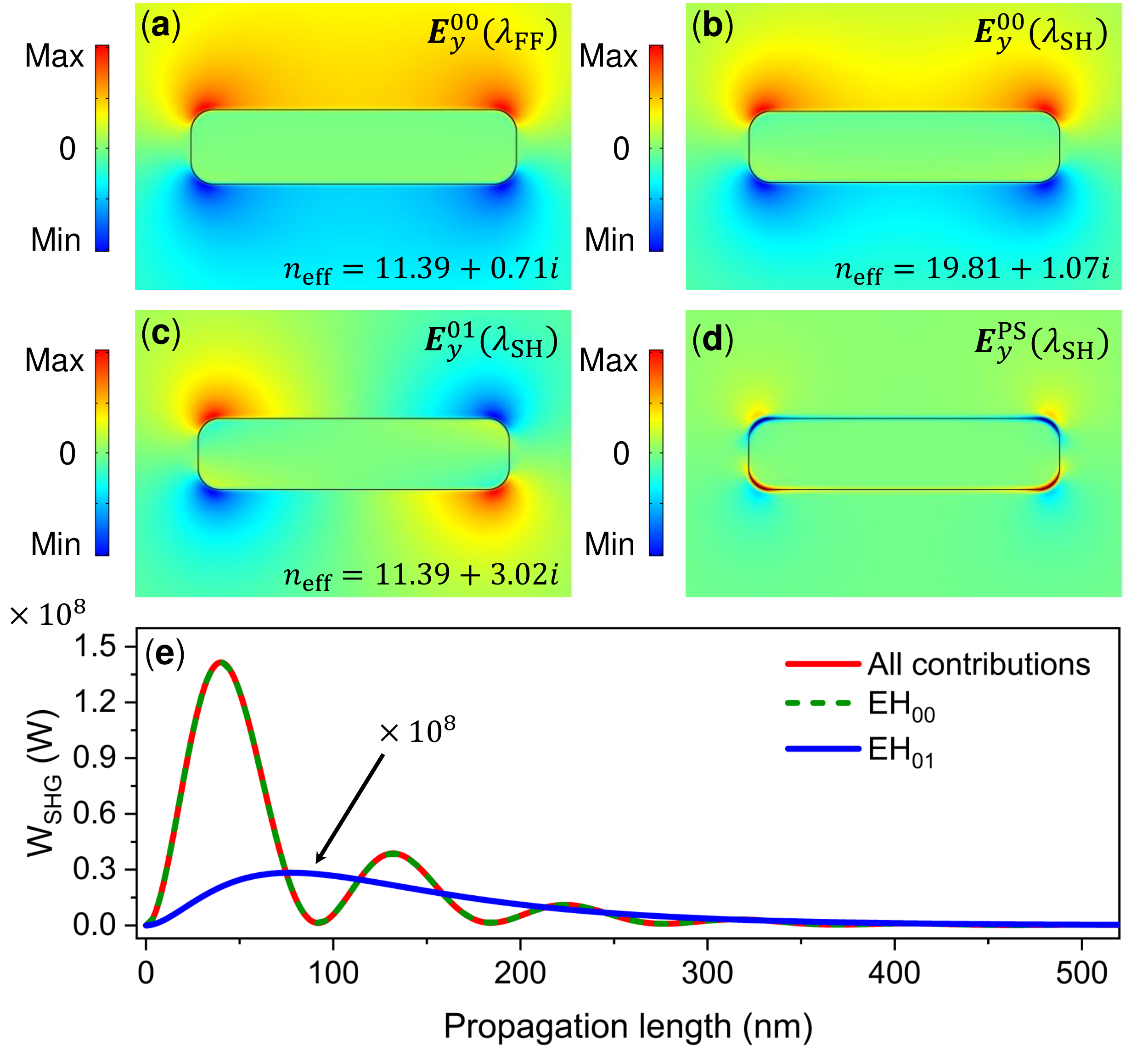}
  \end{center}
\caption{\small Electric field profile ($E_y$\textendash component) of the (a) fundamental mode ${\rm EH_{00}@\lambda_{FF}}$, (b) first mode ${\rm EH_{00}@\lambda_{SH}}$, (c) the second mode ${\rm EH_{01}@\lambda_{SH}}$ at the SH wavelength, and (d) the particular solution (PS). (e) SHG intensity as a function of propagation distance along with the individual contribution of the each mode. The propagation constant of the ${\rm EH_{00}@\lambda_{FF}}$ is $\kappa_{\rm FF}= 4.62\times 10^7+2.88\times 10^7i$ whereas those of ${\rm EH_{00}@\lambda_{SH}}$ and ${\rm EH_{01}@\lambda_{SH}}$ are $\kappa_{\rm EH00}=1.6\times 10^8+8.7\times 10^6i$ and $\kappa_{\rm EH01}=9.23\times 10^7+2.44\times 10^7i$, respectively.}
  \label{Fig:SO_Fig2}
\end{figure}

In order to show an example of the proposed formulation with electron spill-out within the framework of QHT, we study SHG in a metal strip waveguide of width $w$ and height $h$ immersed in a dielectric medium with a dielectric constant $\varepsilon_d$, as depicted in \crefformat{figure}{Fig.~#2#1{(a)}#3}\cref{Fig:SO_Fig1}.
We compute the space-dependent equilibrium electron density $n_0(\mathbf{r})$ self-consistently using the zero-\textit{th} order QHT equation (see Refs. \cite{Ciraci:2016il, Khalid2021prb} for more details).
The color map and line plot of $n_0$, showing the electron spill-out from the metal-dielectric interface, are presented in \crefformat{figure}{Fig.~#2#1{(b)}#3}\cref{Fig:SO_Fig1} and \crefformat{figure}{~#2#1{(c)}#3}\cref{Fig:SO_Fig1}, respectively.
Considering a fixed waveguide height $h=5$~nm, this configuration supports the hybrid mode ${\rm EH_{00}@\lambda_{FF}}$ at a pump wavelength $\lambda_{\rm FF}=1550$ nm and two hybrid modes ${\rm EH_{00}@\lambda_{SH}}$ and ${\rm EH_{01}@\lambda_{SH}}$ at the SH wavelength $\lambda_{SH}=775$ nm.
The real part of the effective indices of these modes as a function of waveguide width $w$ are plotted in \crefformat{figure}{~#2#1{(d)}#3}\cref{Fig:SO_Fig1}.

The PM between the symmetric mode ${\rm EH_{00}@\lambda_{FF}}$ and the anti-symmetric mode ${\rm EH_{01}@\lambda_{SH}}$ occurs for the waveguide width $w=21.85$~nm.
The associated mode profiles ($E_y$ \textendash component) at the FF and SH are depicted in \crefformat{figure}{Fig.~#2#1{(a-c)}#3}\cref{Fig:SO_Fig2} and the field profile of the particular solution (PS) is shown in \crefformat{figure}{Fig.~#2#1{(d)}#3}\cref{Fig:SO_Fig2}.
To explore the contributions from each mode at the SH to the generated signal, it can be noted that nonlinear source field, i.e. the particular solution, see \crefformat{figure}{Fig.~#2#1{(d)}#3}\cref{Fig:SO_Fig2}, overlaps well with the symmetric mode ${\rm EH_{00}@\lambda_{SH}}$ and, therefore, major contribution to the generated power comes from this mode, as shown in \crefformat{figure}{Fig.~#2#1{(e)}#3}\cref{Fig:SO_Fig2}.
indeed, we can observe that there is no overlap between the nonlinear source (PS) and the ${\rm EH_{01}@\lambda_{SH}}$ mode due to its anti-symmetric nature, resulting in virtually zero contribution to the SHG from this mode.
From this example, we can appreciate how important is to have access to the exact SHG along the waveguide.
In fact, a traditional optimization technique, i.e. the PM technique, might not always provide the most efficient design. 

\section{Conclusions}
We have derived and employed a method to study SHG originating from FE hydrodynamic nonlinearities in plasmonic waveguides. Our technique distinguishes itself from conventional approaches, which often neglect electron pressure effects and other quantum hydrodynamic corrections to surface nonlinear contributions.
Indeed, such elements play a pivotal role in nonlinear interactions, as shown in \cite{Ciraci:2012vw,Khalid:2020bx}.
Moreover, the numerical nature of the PSM allows to easily calculate the response of arbitrary nonlinear sources providing a valuable and flexible tool for nonlinear guided optics. 
In particular, our formalism can be applied to explore FE nonlinearities in mid-IR plasmonic waveguides made of heavily doped semiconductors \cite{Soref:12, gamal:15, Biagioni:15_group, Chang:12, Ramirez:18, Mu:14, Gallacher:18}, which recently emerged as promising high-quality and tunable plasmonic materials in this range of wavelenghts, with many potential applications in IR detection, sensing, optoelectronics and light harvesting \cite{Taliercio:2019ib}. Indeed, although FE optical nonlinearities have mostly been observed in metals, analogous effects may also occur in heavily doped semiconductors and, when coupled with plasmonic enhancement, these nonlinearities could be up to two orders of magnitude larger than conventional semiconductor nonlinearities \cite{deluca2021,deluca:epj2022}.  
%\begin{acknowledgments}
%We wish to acknowledge the support of the author community in using
%REV\TeX{}, offering suggestions and encouragement, testing new versions, \dots.
%\end{acknowledgments}



\begin{thebibliography}{77}%
\makeatletter
\providecommand \@ifxundefined [1]{%
 \@ifx{#1\undefined}
}%
\providecommand \@ifnum [1]{%
 \ifnum #1\expandafter \@firstoftwo
 \else \expandafter \@secondoftwo
 \fi
}%
\providecommand \@ifx [1]{%
 \ifx #1\expandafter \@firstoftwo
 \else \expandafter \@secondoftwo
 \fi
}%
\providecommand \natexlab [1]{#1}%
\providecommand \enquote  [1]{``#1''}%
\providecommand \bibnamefont  [1]{#1}%
\providecommand \bibfnamefont [1]{#1}%
\providecommand \citenamefont [1]{#1}%
\providecommand \href@noop [0]{\@secondoftwo}%
\providecommand \href [0]{\begingroup \@sanitize@url \@href}%
\providecommand \@href[1]{\@@startlink{#1}\@@href}%
\providecommand \@@href[1]{\endgroup#1\@@endlink}%
\providecommand \@sanitize@url [0]{\catcode `\\12\catcode `\$12\catcode
  `\&12\catcode `\#12\catcode `\^12\catcode `\_12\catcode `\%12\relax}%
\providecommand \@@startlink[1]{}%
\providecommand \@@endlink[0]{}%
\providecommand \url  [0]{\begingroup\@sanitize@url \@url }%
\providecommand \@url [1]{\endgroup\@href {#1}{\urlprefix }}%
\providecommand \urlprefix  [0]{URL }%
\providecommand \Eprint [0]{\href }%
\providecommand \doibase [0]{https://doi.org/}%
\providecommand \selectlanguage [0]{\@gobble}%
\providecommand \bibinfo  [0]{\@secondoftwo}%
\providecommand \bibfield  [0]{\@secondoftwo}%
\providecommand \translation [1]{[#1]}%
\providecommand \BibitemOpen [0]{}%
\providecommand \bibitemStop [0]{}%
\providecommand \bibitemNoStop [0]{.\EOS\space}%
\providecommand \EOS [0]{\spacefactor3000\relax}%
\providecommand \BibitemShut  [1]{\csname bibitem#1\endcsname}%
\let\auto@bib@innerbib\@empty
%</preamble>
\bibitem [{\citenamefont {Stockman}\ \emph {et~al.}(2018)\citenamefont
  {Stockman}, \citenamefont {Kneipp}, \citenamefont {Bozhevolnyi},
  \citenamefont {Saha}, \citenamefont {Dutta}, \citenamefont {Ndukaife},
  \citenamefont {Kinsey}, \citenamefont {Reddy}, \citenamefont {Guler},
  \citenamefont {Shalaev} \emph {et~al.}}]{Stockman_2018}%
  \BibitemOpen
  \bibfield  {author} {\bibinfo {author} {\bibfnamefont {M.~I.}\ \bibnamefont
  {Stockman}}, \bibinfo {author} {\bibfnamefont {K.}~\bibnamefont {Kneipp}},
  \bibinfo {author} {\bibfnamefont {S.~I.}\ \bibnamefont {Bozhevolnyi}},
  \bibinfo {author} {\bibfnamefont {S.}~\bibnamefont {Saha}}, \bibinfo {author}
  {\bibfnamefont {A.}~\bibnamefont {Dutta}}, \bibinfo {author} {\bibfnamefont
  {J.}~\bibnamefont {Ndukaife}}, \bibinfo {author} {\bibfnamefont
  {N.}~\bibnamefont {Kinsey}}, \bibinfo {author} {\bibfnamefont
  {H.}~\bibnamefont {Reddy}}, \bibinfo {author} {\bibfnamefont
  {U.}~\bibnamefont {Guler}}, \bibinfo {author} {\bibfnamefont {V.~M.}\
  \bibnamefont {Shalaev}}, \emph {et~al.},\ }\bibfield  {title} {\bibinfo
  {title} {Roadmap on plasmonics},\ }\href
  {https://doi.org/10.1088/2040-8986/aaa114} {\bibfield  {journal} {\bibinfo
  {journal} {J. Optics}\ }\textbf {\bibinfo {volume} {20}},\ \bibinfo {pages}
  {043001} (\bibinfo {year} {2018})}\BibitemShut {NoStop}%
\bibitem [{\citenamefont {Stockman}(2011)}]{stockman2011}%
  \BibitemOpen
  \bibfield  {author} {\bibinfo {author} {\bibfnamefont {M.~I.}\ \bibnamefont
  {Stockman}},\ }\bibfield  {title} {\bibinfo {title} {Nanoplasmonics: past,
  present, and glimpse into future},\ }\href
  {https://doi.org/10.1364/OE.19.022029} {\bibfield  {journal} {\bibinfo
  {journal} {Opt. Express}\ }\textbf {\bibinfo {volume} {19}},\ \bibinfo
  {pages} {22029} (\bibinfo {year} {2011})}\BibitemShut {NoStop}%
\bibitem [{\citenamefont {Anker}\ \emph {et~al.}(2008)\citenamefont {Anker},
  \citenamefont {Hall}, \citenamefont {Lyandres}, \citenamefont {Shah},
  \citenamefont {Zhao},\ and\ \citenamefont {Van~Duyne}}]{anker2010biosensing}%
  \BibitemOpen
  \bibfield  {author} {\bibinfo {author} {\bibfnamefont {J.~N.}\ \bibnamefont
  {Anker}}, \bibinfo {author} {\bibfnamefont {W.~P.}\ \bibnamefont {Hall}},
  \bibinfo {author} {\bibfnamefont {O.}~\bibnamefont {Lyandres}}, \bibinfo
  {author} {\bibfnamefont {N.~C.}\ \bibnamefont {Shah}}, \bibinfo {author}
  {\bibfnamefont {J.}~\bibnamefont {Zhao}},\ and\ \bibinfo {author}
  {\bibfnamefont {R.~P.}\ \bibnamefont {Van~Duyne}},\ }\bibfield  {title}
  {\bibinfo {title} {Biosensing with plasmonic nanosensors},\ }\href
  {https://doi.org/https://doi.org/10.1038/nmat2162} {\bibfield  {journal}
  {\bibinfo  {journal} {Nat. Mater.}\ }\textbf {\bibinfo {volume} {7}},\
  \bibinfo {pages} {442} (\bibinfo {year} {2008})}\BibitemShut {NoStop}%
\bibitem [{\citenamefont {Xu}\ \emph {et~al.}(1999)\citenamefont {Xu},
  \citenamefont {Bjerneld}, \citenamefont {K\"all},\ and\ \citenamefont
  {B\"orjesson}}]{PhysRevLett.83.4357}%
  \BibitemOpen
  \bibfield  {author} {\bibinfo {author} {\bibfnamefont {H.}~\bibnamefont
  {Xu}}, \bibinfo {author} {\bibfnamefont {E.~J.}\ \bibnamefont {Bjerneld}},
  \bibinfo {author} {\bibfnamefont {M.}~\bibnamefont {K\"all}},\ and\ \bibinfo
  {author} {\bibfnamefont {L.}~\bibnamefont {B\"orjesson}},\ }\bibfield
  {title} {\bibinfo {title} {Spectroscopy of single hemoglobin molecules by
  surface enhanced raman scattering},\ }\href
  {https://doi.org/10.1103/PhysRevLett.83.4357} {\bibfield  {journal} {\bibinfo
   {journal} {Phys. Rev. Lett.}\ }\textbf {\bibinfo {volume} {83}},\ \bibinfo
  {pages} {4357} (\bibinfo {year} {1999})}\BibitemShut {NoStop}%
\bibitem [{\citenamefont {Atwater}\ and\ \citenamefont
  {Polman}(2010)}]{atwater2011plasmonics}%
  \BibitemOpen
  \bibfield  {author} {\bibinfo {author} {\bibfnamefont {H.~A.}\ \bibnamefont
  {Atwater}}\ and\ \bibinfo {author} {\bibfnamefont {A.}~\bibnamefont
  {Polman}},\ }\bibfield  {title} {\bibinfo {title} {Plasmonics for improved
  photovoltaic devices},\ }\href
  {https://doi.org/https://doi.org/10.1038/nmat2629} {\bibfield  {journal}
  {\bibinfo  {journal} {Nat. Mater.}\ }\textbf {\bibinfo {volume} {9}},\
  \bibinfo {pages} {205} (\bibinfo {year} {2010})}\BibitemShut {NoStop}%
\bibitem [{\citenamefont {MacDonald}\ \emph {et~al.}(2009)\citenamefont
  {MacDonald}, \citenamefont {S{\'a}mson}, \citenamefont {Stockman},\ and\
  \citenamefont {Zheludev}}]{macdonald2009ultrafast}%
  \BibitemOpen
  \bibfield  {author} {\bibinfo {author} {\bibfnamefont {K.~F.}\ \bibnamefont
  {MacDonald}}, \bibinfo {author} {\bibfnamefont {Z.~L.}\ \bibnamefont
  {S{\'a}mson}}, \bibinfo {author} {\bibfnamefont {M.~I.}\ \bibnamefont
  {Stockman}},\ and\ \bibinfo {author} {\bibfnamefont {N.~I.}\ \bibnamefont
  {Zheludev}},\ }\bibfield  {title} {\bibinfo {title} {Ultrafast active
  plasmonics},\ }\href
  {https://doi.org/https://doi.org/10.1038/nphoton.2008.249} {\bibfield
  {journal} {\bibinfo  {journal} {Nat. Photonics}\ }\textbf {\bibinfo {volume}
  {3}},\ \bibinfo {pages} {55} (\bibinfo {year} {2009})}\BibitemShut {NoStop}%
\bibitem [{\citenamefont {Habib}\ \emph {et~al.}(2020)\citenamefont {Habib},
  \citenamefont {Zhu}, \citenamefont {Fong},\ and\ \citenamefont
  {Yanik}}]{HabibZhuFongYanik+2020+3805+3829}%
  \BibitemOpen
  \bibfield  {author} {\bibinfo {author} {\bibfnamefont {A.}~\bibnamefont
  {Habib}}, \bibinfo {author} {\bibfnamefont {X.}~\bibnamefont {Zhu}}, \bibinfo
  {author} {\bibfnamefont {S.}~\bibnamefont {Fong}},\ and\ \bibinfo {author}
  {\bibfnamefont {A.~A.}\ \bibnamefont {Yanik}},\ }\bibfield  {title} {\bibinfo
  {title} {Active plasmonic nanoantenna: an emerging toolbox from photonics to
  neuroscience},\ }\href {https://doi.org/doi:10.1515/nanoph-2020-0275}
  {\bibfield  {journal} {\bibinfo  {journal} {Nanophotonics}\ }\textbf
  {\bibinfo {volume} {9}},\ \bibinfo {pages} {3805} (\bibinfo {year}
  {2020})}\BibitemShut {NoStop}%
\bibitem [{\citenamefont {Johnson}\ \emph {et~al.}(2002)\citenamefont
  {Johnson}, \citenamefont {Choi}, \citenamefont {Knutsen}, \citenamefont
  {Schaller}, \citenamefont {Yang},\ and\ \citenamefont
  {Saykally}}]{johnson2002single}%
  \BibitemOpen
  \bibfield  {author} {\bibinfo {author} {\bibfnamefont {J.~C.}\ \bibnamefont
  {Johnson}}, \bibinfo {author} {\bibfnamefont {H.-J.}\ \bibnamefont {Choi}},
  \bibinfo {author} {\bibfnamefont {K.~P.}\ \bibnamefont {Knutsen}}, \bibinfo
  {author} {\bibfnamefont {R.~D.}\ \bibnamefont {Schaller}}, \bibinfo {author}
  {\bibfnamefont {P.}~\bibnamefont {Yang}},\ and\ \bibinfo {author}
  {\bibfnamefont {R.~J.}\ \bibnamefont {Saykally}},\ }\bibfield  {title}
  {\bibinfo {title} {Single gallium nitride nanowire lasers},\ }\href
  {https://doi.org/https://doi.org/10.1038/nmat728} {\bibfield  {journal}
  {\bibinfo  {journal} {Nat. Mater.}\ }\textbf {\bibinfo {volume} {1}},\
  \bibinfo {pages} {106} (\bibinfo {year} {2002})}\BibitemShut {NoStop}%
\bibitem [{\citenamefont {Duan}\ \emph {et~al.}(2003)\citenamefont {Duan},
  \citenamefont {Huang}, \citenamefont {Agarwal},\ and\ \citenamefont
  {Lieber}}]{duan2003single}%
  \BibitemOpen
  \bibfield  {author} {\bibinfo {author} {\bibfnamefont {X.}~\bibnamefont
  {Duan}}, \bibinfo {author} {\bibfnamefont {Y.}~\bibnamefont {Huang}},
  \bibinfo {author} {\bibfnamefont {R.}~\bibnamefont {Agarwal}},\ and\ \bibinfo
  {author} {\bibfnamefont {C.~M.}\ \bibnamefont {Lieber}},\ }\bibfield  {title}
  {\bibinfo {title} {Single-nanowire electrically driven lasers},\ }\href
  {https://doi.org/https://doi.org/10.1038/nature01353} {\bibfield  {journal}
  {\bibinfo  {journal} {Nature}\ }\textbf {\bibinfo {volume} {421}},\ \bibinfo
  {pages} {241} (\bibinfo {year} {2003})}\BibitemShut {NoStop}%
\bibitem [{\citenamefont {Oulton}\ \emph {et~al.}(2009)\citenamefont {Oulton},
  \citenamefont {Sorger}, \citenamefont {Zentgraf}, \citenamefont {Ma},
  \citenamefont {Gladden}, \citenamefont {Dai}, \citenamefont {Bartal},\ and\
  \citenamefont {Zhang}}]{oulton2009plasmon}%
  \BibitemOpen
  \bibfield  {author} {\bibinfo {author} {\bibfnamefont {R.~F.}\ \bibnamefont
  {Oulton}}, \bibinfo {author} {\bibfnamefont {V.~J.}\ \bibnamefont {Sorger}},
  \bibinfo {author} {\bibfnamefont {T.}~\bibnamefont {Zentgraf}}, \bibinfo
  {author} {\bibfnamefont {R.-M.}\ \bibnamefont {Ma}}, \bibinfo {author}
  {\bibfnamefont {C.}~\bibnamefont {Gladden}}, \bibinfo {author} {\bibfnamefont
  {L.}~\bibnamefont {Dai}}, \bibinfo {author} {\bibfnamefont {G.}~\bibnamefont
  {Bartal}},\ and\ \bibinfo {author} {\bibfnamefont {X.}~\bibnamefont
  {Zhang}},\ }\bibfield  {title} {\bibinfo {title} {Plasmon lasers at deep
  subwavelength scale},\ }\href
  {https://doi.org/https://doi.org/10.1038/nature08364} {\bibfield  {journal}
  {\bibinfo  {journal} {Nature}\ }\textbf {\bibinfo {volume} {461}},\ \bibinfo
  {pages} {629} (\bibinfo {year} {2009})}\BibitemShut {NoStop}%
\bibitem [{\citenamefont {Zia}\ \emph {et~al.}(2006)\citenamefont {Zia},
  \citenamefont {Schuller}, \citenamefont {Chandran},\ and\ \citenamefont
  {Brongersma}}]{ZIA200620}%
  \BibitemOpen
  \bibfield  {author} {\bibinfo {author} {\bibfnamefont {R.}~\bibnamefont
  {Zia}}, \bibinfo {author} {\bibfnamefont {J.~A.}\ \bibnamefont {Schuller}},
  \bibinfo {author} {\bibfnamefont {A.}~\bibnamefont {Chandran}},\ and\
  \bibinfo {author} {\bibfnamefont {M.~L.}\ \bibnamefont {Brongersma}},\
  }\bibfield  {title} {\bibinfo {title} {Plasmonics: the next chip-scale
  technology},\ }\href
  {https://doi.org/https://doi.org/10.1016/S1369-7021(06)71572-3} {\bibfield
  {journal} {\bibinfo  {journal} {Mater. Today}\ }\textbf {\bibinfo {volume}
  {9}},\ \bibinfo {pages} {20} (\bibinfo {year} {2006})}\BibitemShut {NoStop}%
\bibitem [{\citenamefont {Yang}\ and\ \citenamefont
  {Lu}(2012)}]{yang2012subwavelength}%
  \BibitemOpen
  \bibfield  {author} {\bibinfo {author} {\bibfnamefont {R.}~\bibnamefont
  {Yang}}\ and\ \bibinfo {author} {\bibfnamefont {Z.}~\bibnamefont {Lu}},\
  }\bibfield  {title} {\bibinfo {title} {Subwavelength plasmonic waveguides and
  plasmonic materials},\ }\href {https://doi.org/10.1155/2012/258013}
  {\bibfield  {journal} {\bibinfo  {journal} {Int. J. Opt.}\ }\textbf {\bibinfo
  {volume} {2012}} (\bibinfo {year} {2012})}\BibitemShut {NoStop}%
\bibitem [{\citenamefont {Han}\ and\ \citenamefont
  {Bozhevolnyi}(2012)}]{Han_2012}%
  \BibitemOpen
  \bibfield  {author} {\bibinfo {author} {\bibfnamefont {Z.}~\bibnamefont
  {Han}}\ and\ \bibinfo {author} {\bibfnamefont {S.~I.}\ \bibnamefont
  {Bozhevolnyi}},\ }\bibfield  {title} {\bibinfo {title} {Radiation guiding
  with surface plasmon polaritons},\ }\href
  {https://doi.org/10.1088/0034-4885/76/1/016402} {\bibfield  {journal}
  {\bibinfo  {journal} {Reports on Progress in Physics}\ }\textbf {\bibinfo
  {volume} {76}},\ \bibinfo {pages} {016402} (\bibinfo {year}
  {2012})}\BibitemShut {NoStop}%
\bibitem [{\citenamefont {Fang}\ and\ \citenamefont
  {Sun}(2015)}]{fang2015nanoplasmonic}%
  \BibitemOpen
  \bibfield  {author} {\bibinfo {author} {\bibfnamefont {Y.}~\bibnamefont
  {Fang}}\ and\ \bibinfo {author} {\bibfnamefont {M.}~\bibnamefont {Sun}},\
  }\bibfield  {title} {\bibinfo {title} {Nanoplasmonic waveguides: towards
  applications in integrated nanophotonic circuits},\ }\href
  {https://doi.org/https://doi.org/10.1038/lsa.2015.67} {\bibfield  {journal}
  {\bibinfo  {journal} {Light Sci. Appl.}\ }\textbf {\bibinfo {volume} {4}},\
  \bibinfo {pages} {e294} (\bibinfo {year} {2015})}\BibitemShut {NoStop}%
\bibitem [{\citenamefont {Wei}\ \emph {et~al.}(2011{\natexlab{a}})\citenamefont
  {Wei}, \citenamefont {Li}, \citenamefont {Tian}, \citenamefont {Wang},
  \citenamefont {Cong}, \citenamefont {Liu}, \citenamefont {Zhang},
  \citenamefont {Nordlander}, \citenamefont {Halas},\ and\ \citenamefont
  {Xu}}]{wei2011quantum}%
  \BibitemOpen
  \bibfield  {author} {\bibinfo {author} {\bibfnamefont {H.}~\bibnamefont
  {Wei}}, \bibinfo {author} {\bibfnamefont {Z.}~\bibnamefont {Li}}, \bibinfo
  {author} {\bibfnamefont {X.}~\bibnamefont {Tian}}, \bibinfo {author}
  {\bibfnamefont {Z.}~\bibnamefont {Wang}}, \bibinfo {author} {\bibfnamefont
  {F.}~\bibnamefont {Cong}}, \bibinfo {author} {\bibfnamefont {N.}~\bibnamefont
  {Liu}}, \bibinfo {author} {\bibfnamefont {S.}~\bibnamefont {Zhang}}, \bibinfo
  {author} {\bibfnamefont {P.}~\bibnamefont {Nordlander}}, \bibinfo {author}
  {\bibfnamefont {N.~J.}\ \bibnamefont {Halas}},\ and\ \bibinfo {author}
  {\bibfnamefont {H.}~\bibnamefont {Xu}},\ }\bibfield  {title} {\bibinfo
  {title} {Quantum dot-based local field imaging reveals plasmon-based
  interferometric logic in silver nanowire networks},\ }\href
  {https://doi.org/https://doi.org/10.1021/nl103228b} {\bibfield  {journal}
  {\bibinfo  {journal} {Nano Lett.}\ }\textbf {\bibinfo {volume} {11}},\
  \bibinfo {pages} {471} (\bibinfo {year} {2011}{\natexlab{a}})}\BibitemShut
  {NoStop}%
\bibitem [{\citenamefont {Wei}\ \emph {et~al.}(2011{\natexlab{b}})\citenamefont
  {Wei}, \citenamefont {Wang}, \citenamefont {Tian}, \citenamefont {K{\"a}ll},\
  and\ \citenamefont {Xu}}]{wei2011cascaded}%
  \BibitemOpen
  \bibfield  {author} {\bibinfo {author} {\bibfnamefont {H.}~\bibnamefont
  {Wei}}, \bibinfo {author} {\bibfnamefont {Z.}~\bibnamefont {Wang}}, \bibinfo
  {author} {\bibfnamefont {X.}~\bibnamefont {Tian}}, \bibinfo {author}
  {\bibfnamefont {M.}~\bibnamefont {K{\"a}ll}},\ and\ \bibinfo {author}
  {\bibfnamefont {H.}~\bibnamefont {Xu}},\ }\bibfield  {title} {\bibinfo
  {title} {Cascaded logic gates in nanophotonic plasmon networks},\ }\href
  {https://doi.org/https://doi.org/10.1038/ncomms1388} {\bibfield  {journal}
  {\bibinfo  {journal} {Nature communications}\ }\textbf {\bibinfo {volume}
  {2}},\ \bibinfo {pages} {1} (\bibinfo {year}
  {2011}{\natexlab{b}})}\BibitemShut {NoStop}%
\bibitem [{\citenamefont {Chang}\ \emph {et~al.}(2007)\citenamefont {Chang},
  \citenamefont {S{\o}rensen}, \citenamefont {Demler},\ and\ \citenamefont
  {Lukin}}]{chang2007single}%
  \BibitemOpen
  \bibfield  {author} {\bibinfo {author} {\bibfnamefont {D.~E.}\ \bibnamefont
  {Chang}}, \bibinfo {author} {\bibfnamefont {A.~S.}\ \bibnamefont
  {S{\o}rensen}}, \bibinfo {author} {\bibfnamefont {E.~A.}\ \bibnamefont
  {Demler}},\ and\ \bibinfo {author} {\bibfnamefont {M.~D.}\ \bibnamefont
  {Lukin}},\ }\bibfield  {title} {\bibinfo {title} {A single-photon transistor
  using nanoscale surface plasmons},\ }\href
  {https://doi.org/https://doi.org/10.1038/nphys708} {\bibfield  {journal}
  {\bibinfo  {journal} {Nat. Phys.}\ }\textbf {\bibinfo {volume} {3}},\
  \bibinfo {pages} {807} (\bibinfo {year} {2007})}\BibitemShut {NoStop}%
\bibitem [{\citenamefont {Fang}\ \emph {et~al.}(2010)\citenamefont {Fang},
  \citenamefont {Li}, \citenamefont {Huang}, \citenamefont {Zhang},
  \citenamefont {Nordlander}, \citenamefont {Halas},\ and\ \citenamefont
  {Xu}}]{fang2010branched}%
  \BibitemOpen
  \bibfield  {author} {\bibinfo {author} {\bibfnamefont {Y.}~\bibnamefont
  {Fang}}, \bibinfo {author} {\bibfnamefont {Z.}~\bibnamefont {Li}}, \bibinfo
  {author} {\bibfnamefont {Y.}~\bibnamefont {Huang}}, \bibinfo {author}
  {\bibfnamefont {S.}~\bibnamefont {Zhang}}, \bibinfo {author} {\bibfnamefont
  {P.}~\bibnamefont {Nordlander}}, \bibinfo {author} {\bibfnamefont {N.~J.}\
  \bibnamefont {Halas}},\ and\ \bibinfo {author} {\bibfnamefont
  {H.}~\bibnamefont {Xu}},\ }\bibfield  {title} {\bibinfo {title} {Branched
  silver nanowires as controllable plasmon routers},\ }\href
  {https://doi.org/https://doi.org/10.1021/nl101168u} {\bibfield  {journal}
  {\bibinfo  {journal} {Nano Lett.}\ }\textbf {\bibinfo {volume} {10}},\
  \bibinfo {pages} {1950} (\bibinfo {year} {2010})}\BibitemShut {NoStop}%
\bibitem [{\citenamefont {Shalin}\ \emph {et~al.}(2014)\citenamefont {Shalin},
  \citenamefont {Ginzburg}, \citenamefont {Belov}, \citenamefont {Kivshar},\
  and\ \citenamefont {Zayats}}]{shalin2014nano}%
  \BibitemOpen
  \bibfield  {author} {\bibinfo {author} {\bibfnamefont {A.~S.}\ \bibnamefont
  {Shalin}}, \bibinfo {author} {\bibfnamefont {P.}~\bibnamefont {Ginzburg}},
  \bibinfo {author} {\bibfnamefont {P.~A.}\ \bibnamefont {Belov}}, \bibinfo
  {author} {\bibfnamefont {Y.~S.}\ \bibnamefont {Kivshar}},\ and\ \bibinfo
  {author} {\bibfnamefont {A.~V.}\ \bibnamefont {Zayats}},\ }\bibfield  {title}
  {\bibinfo {title} {Nano-opto-mechanical effects in plasmonic waveguides},\
  }\href {https://doi.org/https://doi.org/10.1002/lpor.201300109} {\bibfield
  {journal} {\bibinfo  {journal} {Laser Photonics Rev.}\ }\textbf {\bibinfo
  {volume} {8}},\ \bibinfo {pages} {131} (\bibinfo {year} {2014})}\BibitemShut
  {NoStop}%
\bibitem [{\citenamefont {Ming}\ \emph {et~al.}(2010)\citenamefont {Ming},
  \citenamefont {Zhao}, \citenamefont {Xiao},\ and\ \citenamefont
  {Wang}}]{ming2010resonance}%
  \BibitemOpen
  \bibfield  {author} {\bibinfo {author} {\bibfnamefont {T.}~\bibnamefont
  {Ming}}, \bibinfo {author} {\bibfnamefont {L.}~\bibnamefont {Zhao}}, \bibinfo
  {author} {\bibfnamefont {M.}~\bibnamefont {Xiao}},\ and\ \bibinfo {author}
  {\bibfnamefont {J.}~\bibnamefont {Wang}},\ }\bibfield  {title} {\bibinfo
  {title} {Resonance-coupling-based plasmonic switches},\ }\href
  {https://doi.org/https://doi.org/10.1002/smll.201000920} {\bibfield
  {journal} {\bibinfo  {journal} {Small}\ }\textbf {\bibinfo {volume} {6}},\
  \bibinfo {pages} {2514} (\bibinfo {year} {2010})}\BibitemShut {NoStop}%
\bibitem [{\citenamefont {Chang}\ \emph {et~al.}(2006)\citenamefont {Chang},
  \citenamefont {S\o{}rensen}, \citenamefont {Hemmer},\ and\ \citenamefont
  {Lukin}}]{PhysRevLett.97.053002}%
  \BibitemOpen
  \bibfield  {author} {\bibinfo {author} {\bibfnamefont {D.~E.}\ \bibnamefont
  {Chang}}, \bibinfo {author} {\bibfnamefont {A.~S.}\ \bibnamefont
  {S\o{}rensen}}, \bibinfo {author} {\bibfnamefont {P.~R.}\ \bibnamefont
  {Hemmer}},\ and\ \bibinfo {author} {\bibfnamefont {M.~D.}\ \bibnamefont
  {Lukin}},\ }\bibfield  {title} {\bibinfo {title} {Quantum optics with surface
  plasmons},\ }\href {https://doi.org/10.1103/PhysRevLett.97.053002} {\bibfield
   {journal} {\bibinfo  {journal} {Phys. Rev. Lett.}\ }\textbf {\bibinfo
  {volume} {97}},\ \bibinfo {pages} {053002} (\bibinfo {year}
  {2006})}\BibitemShut {NoStop}%
\bibitem [{\citenamefont {Huck}\ and\ \citenamefont {Andersen}(2016)}]{Huck}%
  \BibitemOpen
  \bibfield  {author} {\bibinfo {author} {\bibfnamefont {A.}~\bibnamefont
  {Huck}}\ and\ \bibinfo {author} {\bibfnamefont {U.~L.}\ \bibnamefont
  {Andersen}},\ }\bibfield  {title} {\bibinfo {title} {Coupling single emitters
  to quantum plasmonic circuits},\ }\href
  {https://doi.org/doi:10.1515/nanoph-2015-0153} {\bibfield  {journal}
  {\bibinfo  {journal} {Nanophotonics}\ }\textbf {\bibinfo {volume} {5}},\
  \bibinfo {pages} {483} (\bibinfo {year} {2016})}\BibitemShut {NoStop}%
\bibitem [{\citenamefont {Martin-Cano}\ \emph {et~al.}(2010)\citenamefont
  {Martin-Cano}, \citenamefont {Martin-Moreno}, \citenamefont {Garcia-Vidal},\
  and\ \citenamefont {Moreno}}]{martin2010resonance}%
  \BibitemOpen
  \bibfield  {author} {\bibinfo {author} {\bibfnamefont {D.}~\bibnamefont
  {Martin-Cano}}, \bibinfo {author} {\bibfnamefont {L.}~\bibnamefont
  {Martin-Moreno}}, \bibinfo {author} {\bibfnamefont {F.~J.}\ \bibnamefont
  {Garcia-Vidal}},\ and\ \bibinfo {author} {\bibfnamefont {E.}~\bibnamefont
  {Moreno}},\ }\bibfield  {title} {\bibinfo {title} {Resonance energy transfer
  and superradiance mediated by plasmonic nanowaveguides},\ }\href
  {https://doi.org/https://doi.org/10.1021/nl101876f} {\bibfield  {journal}
  {\bibinfo  {journal} {Nano Lett.}\ }\textbf {\bibinfo {volume} {10}},\
  \bibinfo {pages} {3129} (\bibinfo {year} {2010})}\BibitemShut {NoStop}%
\bibitem [{\citenamefont {Gonzalez-Tudela}\ \emph {et~al.}(2011)\citenamefont
  {Gonzalez-Tudela}, \citenamefont {Martin-Cano}, \citenamefont {Moreno},
  \citenamefont {Martin-Moreno}, \citenamefont {Tejedor},\ and\ \citenamefont
  {Garcia-Vidal}}]{PhysRevLett.106.020501}%
  \BibitemOpen
  \bibfield  {author} {\bibinfo {author} {\bibfnamefont {A.}~\bibnamefont
  {Gonzalez-Tudela}}, \bibinfo {author} {\bibfnamefont {D.}~\bibnamefont
  {Martin-Cano}}, \bibinfo {author} {\bibfnamefont {E.}~\bibnamefont {Moreno}},
  \bibinfo {author} {\bibfnamefont {L.}~\bibnamefont {Martin-Moreno}}, \bibinfo
  {author} {\bibfnamefont {C.}~\bibnamefont {Tejedor}},\ and\ \bibinfo {author}
  {\bibfnamefont {F.~J.}\ \bibnamefont {Garcia-Vidal}},\ }\bibfield  {title}
  {\bibinfo {title} {Entanglement of two qubits mediated by one-dimensional
  plasmonic waveguides},\ }\href
  {https://doi.org/10.1103/PhysRevLett.106.020501} {\bibfield  {journal}
  {\bibinfo  {journal} {Phys. Rev. Lett.}\ }\textbf {\bibinfo {volume} {106}},\
  \bibinfo {pages} {020501} (\bibinfo {year} {2011})}\BibitemShut {NoStop}%
\bibitem [{\citenamefont {Klein}\ \emph {et~al.}(2006)\citenamefont {Klein},
  \citenamefont {Enkrich}, \citenamefont {Wegener},\ and\ \citenamefont
  {Linden}}]{klein2006second}%
  \BibitemOpen
  \bibfield  {author} {\bibinfo {author} {\bibfnamefont {M.~W.}\ \bibnamefont
  {Klein}}, \bibinfo {author} {\bibfnamefont {C.}~\bibnamefont {Enkrich}},
  \bibinfo {author} {\bibfnamefont {M.}~\bibnamefont {Wegener}},\ and\ \bibinfo
  {author} {\bibfnamefont {S.}~\bibnamefont {Linden}},\ }\bibfield  {title}
  {\bibinfo {title} {Second-harmonic generation from magnetic metamaterials},\
  }\href {https://doi.org/10.1126/science.1129198} {\bibfield  {journal}
  {\bibinfo  {journal} {Science}\ }\textbf {\bibinfo {volume} {313}},\ \bibinfo
  {pages} {502} (\bibinfo {year} {2006})}\BibitemShut {NoStop}%
\bibitem [{\citenamefont {Zeng}\ \emph {et~al.}(9 06)\citenamefont {Zeng},
  \citenamefont {Hoyer}, \citenamefont {Liu}, \citenamefont {Koch},\ and\
  \citenamefont {Moloney}}]{Zeng:2009hd}%
  \BibitemOpen
  \bibfield  {author} {\bibinfo {author} {\bibfnamefont {Y.}~\bibnamefont
  {Zeng}}, \bibinfo {author} {\bibfnamefont {W.}~\bibnamefont {Hoyer}},
  \bibinfo {author} {\bibfnamefont {J.}~\bibnamefont {Liu}}, \bibinfo {author}
  {\bibfnamefont {S.}~\bibnamefont {Koch}},\ and\ \bibinfo {author}
  {\bibfnamefont {J.}~\bibnamefont {Moloney}},\ }\bibfield  {title} {\bibinfo
  {title} {Classical theory for second-harmonic generation from metallic
  nanoparticles},\ }\href {https://doi.org/10.1103/physrevb.79.235109}
  {\bibfield  {journal} {\bibinfo  {journal} {Physical Review B}\ }\textbf
  {\bibinfo {volume} {79}},\ \bibinfo {pages} {235109} (\bibinfo {year}
  {2009-06})}\BibitemShut {NoStop}%
\bibitem [{\citenamefont {Schuller}\ \emph {et~al.}(2010)\citenamefont
  {Schuller}, \citenamefont {Barnard}, \citenamefont {Cai}, \citenamefont
  {Jun}, \citenamefont {White},\ and\ \citenamefont
  {Brongersma}}]{schuller2010plasmonics}%
  \BibitemOpen
  \bibfield  {author} {\bibinfo {author} {\bibfnamefont {J.~A.}\ \bibnamefont
  {Schuller}}, \bibinfo {author} {\bibfnamefont {E.~S.}\ \bibnamefont
  {Barnard}}, \bibinfo {author} {\bibfnamefont {W.}~\bibnamefont {Cai}},
  \bibinfo {author} {\bibfnamefont {Y.~C.}\ \bibnamefont {Jun}}, \bibinfo
  {author} {\bibfnamefont {J.~S.}\ \bibnamefont {White}},\ and\ \bibinfo
  {author} {\bibfnamefont {M.~L.}\ \bibnamefont {Brongersma}},\ }\bibfield
  {title} {\bibinfo {title} {Plasmonics for extreme light concentration and
  manipulation},\ }\href {https://doi.org/https://doi.org/10.1038/nmat2630}
  {\bibfield  {journal} {\bibinfo  {journal} {Nat. Mater.}\ }\textbf {\bibinfo
  {volume} {9}},\ \bibinfo {pages} {193} (\bibinfo {year} {2010})}\BibitemShut
  {NoStop}%
\bibitem [{\citenamefont {Scalora}\ \emph {et~al.}(2010)\citenamefont
  {Scalora}, \citenamefont {Vincenti}, \citenamefont {Ceglia}, \citenamefont
  {Roppo}, \citenamefont {Centini}, \citenamefont {Akozbek},\ and\
  \citenamefont {Bloemer}}]{Scalora:2010kd}%
  \BibitemOpen
  \bibfield  {author} {\bibinfo {author} {\bibfnamefont {M.}~\bibnamefont
  {Scalora}}, \bibinfo {author} {\bibfnamefont {M.~A.}\ \bibnamefont
  {Vincenti}}, \bibinfo {author} {\bibfnamefont {D.~d.}\ \bibnamefont
  {Ceglia}}, \bibinfo {author} {\bibfnamefont {V.}~\bibnamefont {Roppo}},
  \bibinfo {author} {\bibfnamefont {M.}~\bibnamefont {Centini}}, \bibinfo
  {author} {\bibfnamefont {N.}~\bibnamefont {Akozbek}},\ and\ \bibinfo {author}
  {\bibfnamefont {M.~J.}\ \bibnamefont {Bloemer}},\ }\bibfield  {title}
  {\bibinfo {title} {Second- and third-harmonic generation in metal-based
  structures},\ }\href {https://doi.org/10.1103/physreva.82.043828} {\bibfield
  {journal} {\bibinfo  {journal} {Phys. Rev. A}\ }\textbf {\bibinfo {volume}
  {82}},\ \bibinfo {pages} {043828} (\bibinfo {year} {2010})}\BibitemShut
  {NoStop}%
\bibitem [{\citenamefont {Ciracì}\ \emph {et~al.}(2012)\citenamefont
  {Ciracì}, \citenamefont {Poutrina}, \citenamefont {Scalora},\ and\
  \citenamefont {Smith}}]{Ciraci:2012vw}%
  \BibitemOpen
  \bibfield  {author} {\bibinfo {author} {\bibfnamefont {C.}~\bibnamefont
  {Ciracì}}, \bibinfo {author} {\bibfnamefont {E.}~\bibnamefont {Poutrina}},
  \bibinfo {author} {\bibfnamefont {M.}~\bibnamefont {Scalora}},\ and\ \bibinfo
  {author} {\bibfnamefont {D.~R.}\ \bibnamefont {Smith}},\ }\bibfield  {title}
  {\bibinfo {title} {Second-harmonic generation in metallic nanoparticles:
  Clarification of the role of the surface},\ }\href
  {https://doi.org/10.1103/physrevb.86.115451} {\bibfield  {journal} {\bibinfo
  {journal} {Phy. Rev. B}\ }\textbf {\bibinfo {volume} {86}},\ \bibinfo {pages}
  {115451} (\bibinfo {year} {2012})}\BibitemShut {NoStop}%
\bibitem [{\citenamefont {Kauranen}\ and\ \citenamefont
  {Zayats}(2012)}]{kauranen2012nonlinear}%
  \BibitemOpen
  \bibfield  {author} {\bibinfo {author} {\bibfnamefont {M.}~\bibnamefont
  {Kauranen}}\ and\ \bibinfo {author} {\bibfnamefont {A.~V.}\ \bibnamefont
  {Zayats}},\ }\bibfield  {title} {\bibinfo {title} {Nonlinear plasmonics},\
  }\href {https://doi.org/https://doi.org/10.1038/nphoton.2012.244} {\bibfield
  {journal} {\bibinfo  {journal} {Nature Photonics}\ }\textbf {\bibinfo
  {volume} {6}},\ \bibinfo {pages} {737} (\bibinfo {year} {2012})}\BibitemShut
  {NoStop}%
\bibitem [{\citenamefont {Krasavin}\ \emph {et~al.}(2018)\citenamefont
  {Krasavin}, \citenamefont {Ginzburg},\ and\ \citenamefont
  {Zayats}}]{krasavin2018free}%
  \BibitemOpen
  \bibfield  {author} {\bibinfo {author} {\bibfnamefont {A.~V.}\ \bibnamefont
  {Krasavin}}, \bibinfo {author} {\bibfnamefont {P.}~\bibnamefont {Ginzburg}},\
  and\ \bibinfo {author} {\bibfnamefont {A.~V.}\ \bibnamefont {Zayats}},\
  }\bibfield  {title} {\bibinfo {title} {Free-electron optical nonlinearities
  in plasmonic nanostructures: a review of the hydrodynamic description},\
  }\href {https://doi.org/https://doi.org/10.1002/lpor.201700082} {\bibfield
  {journal} {\bibinfo  {journal} {Laser Photonics Rev.}\ }\textbf {\bibinfo
  {volume} {12}},\ \bibinfo {pages} {1700082} (\bibinfo {year}
  {2018})}\BibitemShut {NoStop}%
\bibitem [{\citenamefont {Bonacina}\ \emph {et~al.}(2020)\citenamefont
  {Bonacina}, \citenamefont {Brevet}, \citenamefont {Finazzi},\ and\
  \citenamefont {Celebrano}}]{bonacina2020harmonic}%
  \BibitemOpen
  \bibfield  {author} {\bibinfo {author} {\bibfnamefont {L.}~\bibnamefont
  {Bonacina}}, \bibinfo {author} {\bibfnamefont {P.-F.}\ \bibnamefont
  {Brevet}}, \bibinfo {author} {\bibfnamefont {M.}~\bibnamefont {Finazzi}},\
  and\ \bibinfo {author} {\bibfnamefont {M.}~\bibnamefont {Celebrano}},\
  }\bibfield  {title} {\bibinfo {title} {Harmonic generation at the
  nanoscale},\ }\href {https://doi.org/https://doi.org/10.1063/5.0006093}
  {\bibfield  {journal} {\bibinfo  {journal} {J. Appl. Phys.}\ }\textbf
  {\bibinfo {volume} {127}},\ \bibinfo {pages} {230901} (\bibinfo {year}
  {2020})}\BibitemShut {NoStop}%
\bibitem [{\citenamefont {Tuniz}(2021)}]{tuniz2021nanoscale}%
  \BibitemOpen
  \bibfield  {author} {\bibinfo {author} {\bibfnamefont {A.}~\bibnamefont
  {Tuniz}},\ }\bibfield  {title} {\bibinfo {title} {Nanoscale nonlinear
  plasmonics in photonic waveguides and circuits},\ }\href
  {https://doi.org/https://doi.org/10.1007/s40766-021-00018-7} {\bibfield
  {journal} {\bibinfo  {journal} {Riv. del Nuovo Cim.}\ ,\ \bibinfo {pages}
  {1}} (\bibinfo {year} {2021})}\BibitemShut {NoStop}%
\bibitem [{\citenamefont {Park}\ \emph {et~al.}(2015)\citenamefont {Park},
  \citenamefont {Lu},\ and\ \citenamefont {Ahn}}]{C5CS00050E}%
  \BibitemOpen
  \bibfield  {author} {\bibinfo {author} {\bibfnamefont {W.}~\bibnamefont
  {Park}}, \bibinfo {author} {\bibfnamefont {D.}~\bibnamefont {Lu}},\ and\
  \bibinfo {author} {\bibfnamefont {S.}~\bibnamefont {Ahn}},\ }\bibfield
  {title} {\bibinfo {title} {Plasmon enhancement of luminescence
  upconversion},\ }\href {https://doi.org/10.1039/C5CS00050E} {\bibfield
  {journal} {\bibinfo  {journal} {Chem. Soc. Rev.}\ }\textbf {\bibinfo {volume}
  {44}},\ \bibinfo {pages} {2940} (\bibinfo {year} {2015})}\BibitemShut
  {NoStop}%
\bibitem [{\citenamefont {De~Luca}\ and\ \citenamefont
  {Cirac\`{i}}(2019)}]{DeLuca:19}%
  \BibitemOpen
  \bibfield  {author} {\bibinfo {author} {\bibfnamefont {F.}~\bibnamefont
  {De~Luca}}\ and\ \bibinfo {author} {\bibfnamefont {C.}~\bibnamefont
  {Cirac\`{i}}},\ }\bibfield  {title} {\bibinfo {title} {Difference-frequency
  generation in plasmonic nanostructures: a parameter-free hydrodynamic
  description},\ }\href {https://doi.org/10.1364/JOSAB.36.001979} {\bibfield
  {journal} {\bibinfo  {journal} {J. Opt. Soc. Am. B}\ }\textbf {\bibinfo
  {volume} {36}},\ \bibinfo {pages} {1979} (\bibinfo {year}
  {2019})}\BibitemShut {NoStop}%
\bibitem [{\citenamefont {Boyd}(2006)}]{Boyd:2006uq}%
  \BibitemOpen
  \bibfield  {author} {\bibinfo {author} {\bibfnamefont {R.~W.}\ \bibnamefont
  {Boyd}},\ }\href@noop {} {\emph {\bibinfo {title} {Nonlinear Optics}}}\
  (\bibinfo  {publisher} {Academic Press, San Diego, CA},\ \bibinfo {year}
  {2006})\BibitemShut {NoStop}%
\bibitem [{\citenamefont {Garmire}(2013)}]{Garmire:13}%
  \BibitemOpen
  \bibfield  {author} {\bibinfo {author} {\bibfnamefont {E.}~\bibnamefont
  {Garmire}},\ }\bibfield  {title} {\bibinfo {title} {Nonlinear optics in daily
  life},\ }\href {https://doi.org/10.1364/OE.21.030532} {\bibfield  {journal}
  {\bibinfo  {journal} {Opt. Express}\ }\textbf {\bibinfo {volume} {21}},\
  \bibinfo {pages} {30532} (\bibinfo {year} {2013})}\BibitemShut {NoStop}%
\bibitem [{\citenamefont {Boardman}\ \emph {et~al.}(2012)\citenamefont
  {Boardman}, \citenamefont {Pavlov},\ and\ \citenamefont
  {Tanev}}]{boardman2012advanced}%
  \BibitemOpen
  \bibfield  {author} {\bibinfo {author} {\bibfnamefont {A.~D.}\ \bibnamefont
  {Boardman}}, \bibinfo {author} {\bibfnamefont {L.}~\bibnamefont {Pavlov}},\
  and\ \bibinfo {author} {\bibfnamefont {S.}~\bibnamefont {Tanev}},\
  }\href@noop {} {\emph {\bibinfo {title} {Advanced photonics with second-order
  optically nonlinear processes}}}\ (\bibinfo  {publisher} {Springer Science \&
  Business Media},\ \bibinfo {year} {2012})\BibitemShut {NoStop}%
\bibitem [{\citenamefont {Ruan}\ \emph {et~al.}(2009)\citenamefont {Ruan},
  \citenamefont {Veronis}, \citenamefont {Vodopyanov}, \citenamefont {Fejer},\
  and\ \citenamefont {Fan}}]{Ruan:09}%
  \BibitemOpen
  \bibfield  {author} {\bibinfo {author} {\bibfnamefont {Z.}~\bibnamefont
  {Ruan}}, \bibinfo {author} {\bibfnamefont {G.}~\bibnamefont {Veronis}},
  \bibinfo {author} {\bibfnamefont {K.~L.}\ \bibnamefont {Vodopyanov}},
  \bibinfo {author} {\bibfnamefont {M.~M.}\ \bibnamefont {Fejer}},\ and\
  \bibinfo {author} {\bibfnamefont {S.}~\bibnamefont {Fan}},\ }\bibfield
  {title} {\bibinfo {title} {Enhancement of optics-to-thz conversion efficiency
  by metallic slot waveguides},\ }\href
  {http://www.opticsexpress.org/abstract.cfm?URI=oe-17-16-13502} {\bibfield
  {journal} {\bibinfo  {journal} {Opt. Express}\ }\textbf {\bibinfo {volume}
  {17}},\ \bibinfo {pages} {13502} (\bibinfo {year} {2009})}\BibitemShut
  {NoStop}%
\bibitem [{\citenamefont {Davoyan}\ \emph {et~al.}(2010)\citenamefont
  {Davoyan}, \citenamefont {Shadrivov}, \citenamefont {Bozhevolnyi},\ and\
  \citenamefont {Kivshar}}]{davoyan2010backward}%
  \BibitemOpen
  \bibfield  {author} {\bibinfo {author} {\bibfnamefont {A.~R.}\ \bibnamefont
  {Davoyan}}, \bibinfo {author} {\bibfnamefont {I.~V.}\ \bibnamefont
  {Shadrivov}}, \bibinfo {author} {\bibfnamefont {S.~I.}\ \bibnamefont
  {Bozhevolnyi}},\ and\ \bibinfo {author} {\bibfnamefont {Y.~S.}\ \bibnamefont
  {Kivshar}},\ }\bibfield  {title} {\bibinfo {title} {Backward and forward
  modes guided by metal-dielectric-metal plasmonic waveguides},\ }\href
  {https://doi.org/https://doi.org/10.1117/1.3437397} {\bibfield  {journal}
  {\bibinfo  {journal} {J. Nanophotonics}\ }\textbf {\bibinfo {volume} {4}},\
  \bibinfo {pages} {043509} (\bibinfo {year} {2010})}\BibitemShut {NoStop}%
\bibitem [{\citenamefont {Zhang}\ \emph
  {et~al.}(2013{\natexlab{a}})\citenamefont {Zhang}, \citenamefont {Cassan},
  \citenamefont {Gao},\ and\ \citenamefont {Zhang}}]{Zhang:13}%
  \BibitemOpen
  \bibfield  {author} {\bibinfo {author} {\bibfnamefont {J.}~\bibnamefont
  {Zhang}}, \bibinfo {author} {\bibfnamefont {E.}~\bibnamefont {Cassan}},
  \bibinfo {author} {\bibfnamefont {D.}~\bibnamefont {Gao}},\ and\ \bibinfo
  {author} {\bibfnamefont {X.}~\bibnamefont {Zhang}},\ }\bibfield  {title}
  {\bibinfo {title} {Highly efficient phase-matched second harmonic generation
  using an asymmetric plasmonic slot waveguide configuration in hybrid
  polymer-silicon photonics},\ }\href {https://doi.org/10.1364/OE.21.014876}
  {\bibfield  {journal} {\bibinfo  {journal} {Opt. Express}\ }\textbf {\bibinfo
  {volume} {21}},\ \bibinfo {pages} {14876} (\bibinfo {year}
  {2013}{\natexlab{a}})}\BibitemShut {NoStop}%
\bibitem [{\citenamefont {Zhang}\ \emph
  {et~al.}(2013{\natexlab{b}})\citenamefont {Zhang}, \citenamefont {Cassan},\
  and\ \citenamefont {Zhang}}]{Zhang:13a}%
  \BibitemOpen
  \bibfield  {author} {\bibinfo {author} {\bibfnamefont {J.}~\bibnamefont
  {Zhang}}, \bibinfo {author} {\bibfnamefont {E.}~\bibnamefont {Cassan}},\ and\
  \bibinfo {author} {\bibfnamefont {X.}~\bibnamefont {Zhang}},\ }\bibfield
  {title} {\bibinfo {title} {Efficient second harmonic generation from
  mid-infrared to near-infrared regions in silicon-organic hybrid plasmonic
  waveguides with small fabrication-error sensitivity and a large bandwidth},\
  }\href {https://doi.org/10.1364/OL.38.002089} {\bibfield  {journal} {\bibinfo
   {journal} {Opt. Lett.}\ }\textbf {\bibinfo {volume} {38}},\ \bibinfo {pages}
  {2089} (\bibinfo {year} {2013}{\natexlab{b}})}\BibitemShut {NoStop}%
\bibitem [{\citenamefont {Wu}\ \emph {et~al.}(2014)\citenamefont {Wu},
  \citenamefont {Sun}, \citenamefont {Shao}, \citenamefont {Shum},\ and\
  \citenamefont {Huang}}]{Wu:14}%
  \BibitemOpen
  \bibfield  {author} {\bibinfo {author} {\bibfnamefont {T.}~\bibnamefont
  {Wu}}, \bibinfo {author} {\bibfnamefont {Y.}~\bibnamefont {Sun}}, \bibinfo
  {author} {\bibfnamefont {X.}~\bibnamefont {Shao}}, \bibinfo {author}
  {\bibfnamefont {P.~P.}\ \bibnamefont {Shum}},\ and\ \bibinfo {author}
  {\bibfnamefont {T.}~\bibnamefont {Huang}},\ }\bibfield  {title} {\bibinfo
  {title} {Efficient phase-matched third harmonic generation in an asymmetric
  plasmonic slot waveguide},\ }\href {https://doi.org/10.1364/OE.22.018612}
  {\bibfield  {journal} {\bibinfo  {journal} {Opt. Express}\ }\textbf {\bibinfo
  {volume} {22}},\ \bibinfo {pages} {18612} (\bibinfo {year}
  {2014})}\BibitemShut {NoStop}%
\bibitem [{\citenamefont {Sun}\ \emph {et~al.}(2015)\citenamefont {Sun},
  \citenamefont {Zheng}, \citenamefont {Cheng}, \citenamefont {Sun},\ and\
  \citenamefont {Qiao}}]{Sun:15}%
  \BibitemOpen
  \bibfield  {author} {\bibinfo {author} {\bibfnamefont {Y.}~\bibnamefont
  {Sun}}, \bibinfo {author} {\bibfnamefont {Z.}~\bibnamefont {Zheng}}, \bibinfo
  {author} {\bibfnamefont {J.}~\bibnamefont {Cheng}}, \bibinfo {author}
  {\bibfnamefont {G.}~\bibnamefont {Sun}},\ and\ \bibinfo {author}
  {\bibfnamefont {G.}~\bibnamefont {Qiao}},\ }\bibfield  {title} {\bibinfo
  {title} {Highly efficient second harmonic generation in hyperbolic
  metamaterial slot waveguides with large phase matching tolerance},\ }\href
  {https://doi.org/10.1364/OE.23.006370} {\bibfield  {journal} {\bibinfo
  {journal} {Opt. Express}\ }\textbf {\bibinfo {volume} {23}},\ \bibinfo
  {pages} {6370} (\bibinfo {year} {2015})}\BibitemShut {NoStop}%
\bibitem [{\citenamefont {Huang}\ \emph {et~al.}(2016)\citenamefont {Huang},
  \citenamefont {Tagne},\ and\ \citenamefont {Fu}}]{Huang:16}%
  \BibitemOpen
  \bibfield  {author} {\bibinfo {author} {\bibfnamefont {T.}~\bibnamefont
  {Huang}}, \bibinfo {author} {\bibfnamefont {P.~M.}\ \bibnamefont {Tagne}},\
  and\ \bibinfo {author} {\bibfnamefont {S.}~\bibnamefont {Fu}},\ }\bibfield
  {title} {\bibinfo {title} {Efficient second harmonic generation in internal
  asymmetric plasmonic slot waveguide},\ }\href
  {https://doi.org/10.1364/OE.24.009706} {\bibfield  {journal} {\bibinfo
  {journal} {Opt. Express}\ }\textbf {\bibinfo {volume} {24}},\ \bibinfo
  {pages} {9706} (\bibinfo {year} {2016})}\BibitemShut {NoStop}%
\bibitem [{\citenamefont {Shi}\ \emph {et~al.}(2019)\citenamefont {Shi},
  \citenamefont {Li}, \citenamefont {Kang}, \citenamefont {He}, \citenamefont
  {Halas}, \citenamefont {Nordlander}, \citenamefont {Zhang},\ and\
  \citenamefont {Xu}}]{shi2019efficient}%
  \BibitemOpen
  \bibfield  {author} {\bibinfo {author} {\bibfnamefont {J.}~\bibnamefont
  {Shi}}, \bibinfo {author} {\bibfnamefont {Y.}~\bibnamefont {Li}}, \bibinfo
  {author} {\bibfnamefont {M.}~\bibnamefont {Kang}}, \bibinfo {author}
  {\bibfnamefont {X.}~\bibnamefont {He}}, \bibinfo {author} {\bibfnamefont
  {N.~J.}\ \bibnamefont {Halas}}, \bibinfo {author} {\bibfnamefont
  {P.}~\bibnamefont {Nordlander}}, \bibinfo {author} {\bibfnamefont
  {S.}~\bibnamefont {Zhang}},\ and\ \bibinfo {author} {\bibfnamefont
  {H.}~\bibnamefont {Xu}},\ }\bibfield  {title} {\bibinfo {title} {Efficient
  second harmonic generation in a hybrid plasmonic waveguide by mode
  interactions},\ }\href {https://doi.org/10.1021/acs.nanolett.9b01004}
  {\bibfield  {journal} {\bibinfo  {journal} {Nano Letters}\ }\textbf {\bibinfo
  {volume} {19}},\ \bibinfo {pages} {3838} (\bibinfo {year}
  {2019})}\BibitemShut {NoStop}%
\bibitem [{\citenamefont {Thyagarajan}\ \emph {et~al.}(2012)\citenamefont
  {Thyagarajan}, \citenamefont {Rivier}, \citenamefont {Lovera},\ and\
  \citenamefont {Martin}}]{Thyagarajan:12}%
  \BibitemOpen
  \bibfield  {author} {\bibinfo {author} {\bibfnamefont {K.}~\bibnamefont
  {Thyagarajan}}, \bibinfo {author} {\bibfnamefont {S.}~\bibnamefont {Rivier}},
  \bibinfo {author} {\bibfnamefont {A.}~\bibnamefont {Lovera}},\ and\ \bibinfo
  {author} {\bibfnamefont {O.~J.}\ \bibnamefont {Martin}},\ }\bibfield  {title}
  {\bibinfo {title} {Enhanced second-harmonic generation from double resonant
  plasmonic antennae},\ }\href {https://doi.org/10.1364/OE.20.012860}
  {\bibfield  {journal} {\bibinfo  {journal} {Opt. Express}\ }\textbf {\bibinfo
  {volume} {20}},\ \bibinfo {pages} {12860} (\bibinfo {year}
  {2012})}\BibitemShut {NoStop}%
\bibitem [{\citenamefont {Ginzburg}\ \emph {et~al.}(2013)\citenamefont
  {Ginzburg}, \citenamefont {Krasavin},\ and\ \citenamefont
  {Zayats}}]{Ginzburg_2013}%
  \BibitemOpen
  \bibfield  {author} {\bibinfo {author} {\bibfnamefont {P.}~\bibnamefont
  {Ginzburg}}, \bibinfo {author} {\bibfnamefont {A.~V.}\ \bibnamefont
  {Krasavin}},\ and\ \bibinfo {author} {\bibfnamefont {A.~V.}\ \bibnamefont
  {Zayats}},\ }\bibfield  {title} {\bibinfo {title} {Cascaded second-order
  surface plasmon solitons due to intrinsic metal nonlinearity},\ }\href
  {https://doi.org/10.1088/1367-2630/15/1/013031} {\bibfield  {journal}
  {\bibinfo  {journal} {New Journal of Physics}\ }\textbf {\bibinfo {volume}
  {15}},\ \bibinfo {pages} {013031} (\bibinfo {year} {2013})}\BibitemShut
  {NoStop}%
\bibitem [{\citenamefont {Bloembergen}\ \emph {et~al.}(1968)\citenamefont
  {Bloembergen}, \citenamefont {Chang}, \citenamefont {Jha},\ and\
  \citenamefont {Lee}}]{PhysRev.174.813}%
  \BibitemOpen
  \bibfield  {author} {\bibinfo {author} {\bibfnamefont {N.}~\bibnamefont
  {Bloembergen}}, \bibinfo {author} {\bibfnamefont {R.~K.}\ \bibnamefont
  {Chang}}, \bibinfo {author} {\bibfnamefont {S.~S.}\ \bibnamefont {Jha}},\
  and\ \bibinfo {author} {\bibfnamefont {C.~H.}\ \bibnamefont {Lee}},\
  }\bibfield  {title} {\bibinfo {title} {Optical second-harmonic generation in
  reflection from media with inversion symmetry},\ }\href
  {https://doi.org/10.1103/PhysRev.174.813} {\bibfield  {journal} {\bibinfo
  {journal} {Phys. Rev.}\ }\textbf {\bibinfo {volume} {174}},\ \bibinfo {pages}
  {813} (\bibinfo {year} {1968})}\BibitemShut {NoStop}%
\bibitem [{\citenamefont {Khalid}\ and\ \citenamefont
  {Ciracì}(2020)}]{Khalid:2020bx}%
  \BibitemOpen
  \bibfield  {author} {\bibinfo {author} {\bibfnamefont {M.}~\bibnamefont
  {Khalid}}\ and\ \bibinfo {author} {\bibfnamefont {C.}~\bibnamefont
  {Ciracì}},\ }\bibfield  {title} {\bibinfo {title} {Enhancing second-harmonic
  generation with electron spill-out at metallic surfaces},\ }\href
  {https://doi.org/10.1038/s42005-020-00477-0} {\bibfield  {journal} {\bibinfo
  {journal} {Commun. Phys.}\ }\textbf {\bibinfo {volume} {3}},\ \bibinfo
  {pages} {214} (\bibinfo {year} {2020})}\BibitemShut {NoStop}%
\bibitem [{\citenamefont {Zeng}\ \emph {et~al.}(2009)\citenamefont {Zeng},
  \citenamefont {Hoyer}, \citenamefont {Liu}, \citenamefont {Koch},\ and\
  \citenamefont {Moloney}}]{PhysRevB.79.235109}%
  \BibitemOpen
  \bibfield  {author} {\bibinfo {author} {\bibfnamefont {Y.}~\bibnamefont
  {Zeng}}, \bibinfo {author} {\bibfnamefont {W.}~\bibnamefont {Hoyer}},
  \bibinfo {author} {\bibfnamefont {J.}~\bibnamefont {Liu}}, \bibinfo {author}
  {\bibfnamefont {S.~W.}\ \bibnamefont {Koch}},\ and\ \bibinfo {author}
  {\bibfnamefont {J.~V.}\ \bibnamefont {Moloney}},\ }\bibfield  {title}
  {\bibinfo {title} {Classical theory for second-harmonic generation from
  metallic nanoparticles},\ }\href {https://doi.org/10.1103/PhysRevB.79.235109}
  {\bibfield  {journal} {\bibinfo  {journal} {Phys. Rev. B}\ }\textbf {\bibinfo
  {volume} {79}},\ \bibinfo {pages} {235109} (\bibinfo {year}
  {2009})}\BibitemShut {NoStop}%
\bibitem [{\citenamefont {Sipe}\ \emph {et~al.}(1980)\citenamefont {Sipe},
  \citenamefont {So}, \citenamefont {Fukui},\ and\ \citenamefont
  {Stegeman}}]{PhysRevB.21.4389}%
  \BibitemOpen
  \bibfield  {author} {\bibinfo {author} {\bibfnamefont {J.~E.}\ \bibnamefont
  {Sipe}}, \bibinfo {author} {\bibfnamefont {V.~C.~Y.}\ \bibnamefont {So}},
  \bibinfo {author} {\bibfnamefont {M.}~\bibnamefont {Fukui}},\ and\ \bibinfo
  {author} {\bibfnamefont {G.~I.}\ \bibnamefont {Stegeman}},\ }\bibfield
  {title} {\bibinfo {title} {Analysis of second-harmonic generation at metal
  surfaces},\ }\href {https://doi.org/10.1103/PhysRevB.21.4389} {\bibfield
  {journal} {\bibinfo  {journal} {Phys. Rev. B}\ }\textbf {\bibinfo {volume}
  {21}},\ \bibinfo {pages} {4389} (\bibinfo {year} {1980})}\BibitemShut
  {NoStop}%
\bibitem [{\citenamefont {De~Luca}\ \emph {et~al.}(2021)\citenamefont
  {De~Luca}, \citenamefont {Ortolani},\ and\ \citenamefont
  {Cirac\`{\i}}}]{deluca2021}%
  \BibitemOpen
  \bibfield  {author} {\bibinfo {author} {\bibfnamefont {F.}~\bibnamefont
  {De~Luca}}, \bibinfo {author} {\bibfnamefont {M.}~\bibnamefont {Ortolani}},\
  and\ \bibinfo {author} {\bibfnamefont {C.}~\bibnamefont {Cirac\`{\i}}},\
  }\bibfield  {title} {\bibinfo {title} {Free electron nonlinearities in
  heavily doped semiconductors plasmonics},\ }\href
  {https://doi.org/10.1103/PhysRevB.103.115305} {\bibfield  {journal} {\bibinfo
   {journal} {Phys. Rev. B}\ }\textbf {\bibinfo {volume} {103}},\ \bibinfo
  {pages} {115305} (\bibinfo {year} {2021})}\BibitemShut {NoStop}%
\bibitem [{\citenamefont {Ciracì}\ and\ \citenamefont
  {Della~Sala}(2016)}]{Ciraci:2016il}%
  \BibitemOpen
  \bibfield  {author} {\bibinfo {author} {\bibfnamefont {C.}~\bibnamefont
  {Ciracì}}\ and\ \bibinfo {author} {\bibfnamefont {F.}~\bibnamefont
  {Della~Sala}},\ }\bibfield  {title} {\bibinfo {title} {Quantum hydrodynamic
  theory for plasmonics: Impact of the electron density tail},\ }\href
  {https://doi.org/10.1103/physrevb.93.205405} {\bibfield  {journal} {\bibinfo
  {journal} {Phys. Rev. B}\ }\textbf {\bibinfo {volume} {93}},\ \bibinfo
  {pages} {205405} (\bibinfo {year} {2016})}\BibitemShut {NoStop}%
\bibitem [{\citenamefont {Raza}\ \emph {et~al.}(2013)\citenamefont {Raza},
  \citenamefont {Christensen}, \citenamefont {Wubs}, \citenamefont
  {Bozhevolnyi},\ and\ \citenamefont {Mortensen}}]{Raza:2013bd}%
  \BibitemOpen
  \bibfield  {author} {\bibinfo {author} {\bibfnamefont {S.}~\bibnamefont
  {Raza}}, \bibinfo {author} {\bibfnamefont {T.}~\bibnamefont {Christensen}},
  \bibinfo {author} {\bibfnamefont {M.}~\bibnamefont {Wubs}}, \bibinfo {author}
  {\bibfnamefont {S.~I.}\ \bibnamefont {Bozhevolnyi}},\ and\ \bibinfo {author}
  {\bibfnamefont {N.~A.}\ \bibnamefont {Mortensen}},\ }\bibfield  {title}
  {\bibinfo {title} {Nonlocal response in thin-film waveguides: Loss versus
  nonlocality and breaking of complementarity},\ }\href
  {https://doi.org/10.1103/physrevb.88.115401} {\bibfield  {journal} {\bibinfo
  {journal} {Phys. Rev. B}\ }\textbf {\bibinfo {volume} {88}},\ \bibinfo
  {pages} {115401} (\bibinfo {year} {2013})}\BibitemShut {NoStop}%
\bibitem [{\citenamefont {Huang}\ \emph {et~al.}(2013)\citenamefont {Huang},
  \citenamefont {Bao},\ and\ \citenamefont {He}}]{Huang:13}%
  \BibitemOpen
  \bibfield  {author} {\bibinfo {author} {\bibfnamefont {Q.}~\bibnamefont
  {Huang}}, \bibinfo {author} {\bibfnamefont {F.}~\bibnamefont {Bao}},\ and\
  \bibinfo {author} {\bibfnamefont {S.}~\bibnamefont {He}},\ }\bibfield
  {title} {\bibinfo {title} {Nonlocal effects in a hybrid plasmonic waveguide
  for nanoscale confinement},\ }\href {https://doi.org/10.1364/OE.21.001430}
  {\bibfield  {journal} {\bibinfo  {journal} {Opt. Express}\ }\textbf {\bibinfo
  {volume} {21}},\ \bibinfo {pages} {1430} (\bibinfo {year}
  {2013})}\BibitemShut {NoStop}%
\bibitem [{\citenamefont {Toscano}\ \emph {et~al.}(2013)\citenamefont
  {Toscano}, \citenamefont {Raza}, \citenamefont {Yan}, \citenamefont
  {Jeppesen},\ and\ \citenamefont {Xiao}}]{Toscano:2013hr}%
  \BibitemOpen
  \bibfield  {author} {\bibinfo {author} {\bibfnamefont {G.}~\bibnamefont
  {Toscano}}, \bibinfo {author} {\bibfnamefont {S.}~\bibnamefont {Raza}},
  \bibinfo {author} {\bibfnamefont {W.}~\bibnamefont {Yan}}, \bibinfo {author}
  {\bibfnamefont {C.}~\bibnamefont {Jeppesen}},\ and\ \bibinfo {author}
  {\bibfnamefont {S.}~\bibnamefont {Xiao}},\ }\bibfield  {title} {\bibinfo
  {title} {Nonlocal response in plasmonic waveguiding with extreme light
  confinement},\ }\href {https://doi.org/10.1515/nanoph-2013-0014} {\bibfield
  {journal} {\bibinfo  {journal} {Nanophotonics}\ }\textbf {\bibinfo {volume}
  {2}},\ \bibinfo {pages} {161} (\bibinfo {year} {2013})}\BibitemShut {NoStop}%
\bibitem [{\citenamefont {Zheng}\ \emph {et~al.}(2019)\citenamefont {Zheng},
  \citenamefont {Kupresak}, \citenamefont {Moshchalkov}, \citenamefont
  {Mittra},\ and\ \citenamefont {Vandenbosch}}]{Zheng.2019}%
  \BibitemOpen
  \bibfield  {author} {\bibinfo {author} {\bibfnamefont {X.}~\bibnamefont
  {Zheng}}, \bibinfo {author} {\bibfnamefont {M.}~\bibnamefont {Kupresak}},
  \bibinfo {author} {\bibfnamefont {V.~V.}\ \bibnamefont {Moshchalkov}},
  \bibinfo {author} {\bibfnamefont {R.}~\bibnamefont {Mittra}},\ and\ \bibinfo
  {author} {\bibfnamefont {G.~A.~E.}\ \bibnamefont {Vandenbosch}},\ }\bibfield
  {title} {\bibinfo {title} {A potential-based formalism for modeling local and
  hydrodynamic nonlocal responses from plasmonic waveguides},\ }\href
  {https://doi.org/10.1109/tap.2019.2907807} {\bibfield  {journal} {\bibinfo
  {journal} {IEEE Trans. Antennas Propag.}\ }\textbf {\bibinfo {volume} {67}},\
  \bibinfo {pages} {3948} (\bibinfo {year} {2019})}\BibitemShut {NoStop}%
\bibitem [{com()}]{comsol}%
  \BibitemOpen
  \href {http://www.comsol.com} {\bibinfo {title} {Comsol
  multiphysics}}\BibitemShut {NoStop}%
\bibitem [{\citenamefont {McIsaac}(1991)}]{mcisaac1991mode}%
  \BibitemOpen
  \bibfield  {author} {\bibinfo {author} {\bibfnamefont {P.~R.}\ \bibnamefont
  {McIsaac}},\ }\bibfield  {title} {\bibinfo {title} {Mode orthogonality in
  reciprocal and nonreciprocal waveguides},\ }\href@noop {} {\bibfield
  {journal} {\bibinfo  {journal} {IEEE transactions on microwave theory and
  techniques}\ }\textbf {\bibinfo {volume} {39}},\ \bibinfo {pages} {1808}
  (\bibinfo {year} {1991})}\BibitemShut {NoStop}%
\bibitem [{\citenamefont {Mahmoud}(1991)}]{mahmoud1991electromagnetic}%
  \BibitemOpen
  \bibfield  {author} {\bibinfo {author} {\bibfnamefont {S.~F.}\ \bibnamefont
  {Mahmoud}},\ }\href@noop {} {\emph {\bibinfo {title} {Electromagnetic
  waveguides: theory and applications}}},\ \bibinfo {number} {32}\ (\bibinfo
  {publisher} {IET},\ \bibinfo {year} {1991})\BibitemShut {NoStop}%
\bibitem [{\citenamefont {Vidal-Codina}\ \emph {et~al.}(2021)\citenamefont
  {Vidal-Codina}, \citenamefont {Nguyen}, \citenamefont {Ciracì},
  \citenamefont {Oh},\ and\ \citenamefont {Peraire}}]{Vidal-Codina.2021}%
  \BibitemOpen
  \bibfield  {author} {\bibinfo {author} {\bibfnamefont {F.}~\bibnamefont
  {Vidal-Codina}}, \bibinfo {author} {\bibfnamefont {N.-C.}\ \bibnamefont
  {Nguyen}}, \bibinfo {author} {\bibfnamefont {C.}~\bibnamefont {Ciracì}},
  \bibinfo {author} {\bibfnamefont {S.-H.}\ \bibnamefont {Oh}},\ and\ \bibinfo
  {author} {\bibfnamefont {J.}~\bibnamefont {Peraire}},\ }\bibfield  {title}
  {\bibinfo {title} {A nested hybridizable discontinuous galerkin method for
  computing second-harmonic generation in three-dimensional metallic
  nanostructures},\ }\href {https://doi.org/10.1016/j.jcp.2020.110000}
  {\bibfield  {journal} {\bibinfo  {journal} {J. Comput. Phys.}\ }\textbf
  {\bibinfo {volume} {429}},\ \bibinfo {pages} {110000} (\bibinfo {year}
  {2021})}\BibitemShut {NoStop}%
\bibitem [{\citenamefont {Maier}(2007)}]{maier2007plasmonics}%
  \BibitemOpen
  \bibfield  {author} {\bibinfo {author} {\bibfnamefont {S.~A.}\ \bibnamefont
  {Maier}},\ }\href@noop {} {\emph {\bibinfo {title} {Plasmonics: fundamentals
  and applications}}}\ (\bibinfo  {publisher} {Springer Science \& Business
  Media},\ \bibinfo {year} {2007})\BibitemShut {NoStop}%
\bibitem [{\citenamefont {Economou}(1969)}]{economou1969surface}%
  \BibitemOpen
  \bibfield  {author} {\bibinfo {author} {\bibfnamefont {E.}~\bibnamefont
  {Economou}},\ }\bibfield  {title} {\bibinfo {title} {Surface plasmons in thin
  films},\ }\href {https://doi.org/10.1103/PhysRev.182.539} {\bibfield
  {journal} {\bibinfo  {journal} {Phys. Rev.}\ }\textbf {\bibinfo {volume}
  {182}},\ \bibinfo {pages} {539} (\bibinfo {year} {1969})}\BibitemShut
  {NoStop}%
\bibitem [{\citenamefont {Burke}\ \emph {et~al.}(1986)\citenamefont {Burke},
  \citenamefont {Stegeman},\ and\ \citenamefont {Tamir}}]{bruke}%
  \BibitemOpen
  \bibfield  {author} {\bibinfo {author} {\bibfnamefont {J.~J.}\ \bibnamefont
  {Burke}}, \bibinfo {author} {\bibfnamefont {G.~I.}\ \bibnamefont
  {Stegeman}},\ and\ \bibinfo {author} {\bibfnamefont {T.}~\bibnamefont
  {Tamir}},\ }\bibfield  {title} {\bibinfo {title} {Surface-polariton-like
  waves guided by thin, lossy metal films},\ }\href
  {https://doi.org/10.1103/PhysRevB.33.5186} {\bibfield  {journal} {\bibinfo
  {journal} {Phys. Rev. B}\ }\textbf {\bibinfo {volume} {33}},\ \bibinfo
  {pages} {5186} (\bibinfo {year} {1986})}\BibitemShut {NoStop}%
\bibitem [{\citenamefont {Prade}\ \emph {et~al.}(1991)\citenamefont {Prade},
  \citenamefont {Vinet},\ and\ \citenamefont {Mysyrowicz}}]{prade}%
  \BibitemOpen
  \bibfield  {author} {\bibinfo {author} {\bibfnamefont {B.}~\bibnamefont
  {Prade}}, \bibinfo {author} {\bibfnamefont {J.~Y.}\ \bibnamefont {Vinet}},\
  and\ \bibinfo {author} {\bibfnamefont {A.}~\bibnamefont {Mysyrowicz}},\
  }\bibfield  {title} {\bibinfo {title} {Guided optical waves in planar
  heterostructures with negative dielectric constant},\ }\href
  {https://doi.org/10.1103/PhysRevB.44.13556} {\bibfield  {journal} {\bibinfo
  {journal} {Phys. Rev. B}\ }\textbf {\bibinfo {volume} {44}},\ \bibinfo
  {pages} {13556} (\bibinfo {year} {1991})}\BibitemShut {NoStop}%
\bibitem [{\citenamefont {Davoyan}\ \emph {et~al.}(2009)\citenamefont
  {Davoyan}, \citenamefont {Shadrivov},\ and\ \citenamefont
  {Kivshar}}]{Davoyan:09}%
  \BibitemOpen
  \bibfield  {author} {\bibinfo {author} {\bibfnamefont {A.~R.}\ \bibnamefont
  {Davoyan}}, \bibinfo {author} {\bibfnamefont {I.~V.}\ \bibnamefont
  {Shadrivov}},\ and\ \bibinfo {author} {\bibfnamefont {Y.~S.}\ \bibnamefont
  {Kivshar}},\ }\bibfield  {title} {\bibinfo {title} {Quadratic phase matching
  in nonlinear plasmonic nanoscale waveguides},\ }\href
  {https://doi.org/10.1364/OE.17.020063} {\bibfield  {journal} {\bibinfo
  {journal} {Opt. Express}\ }\textbf {\bibinfo {volume} {17}},\ \bibinfo
  {pages} {20063} (\bibinfo {year} {2009})}\BibitemShut {NoStop}%
\bibitem [{\citenamefont {Khalid}\ \emph {et~al.}(2021)\citenamefont {Khalid},
  \citenamefont {Morandi}, \citenamefont {Mallet}, \citenamefont {Hervieux},
  \citenamefont {Manfredi}, \citenamefont {Moreau},\ and\ \citenamefont
  {Cirac\`{\i}}}]{Khalid2021prb}%
  \BibitemOpen
  \bibfield  {author} {\bibinfo {author} {\bibfnamefont {M.}~\bibnamefont
  {Khalid}}, \bibinfo {author} {\bibfnamefont {O.}~\bibnamefont {Morandi}},
  \bibinfo {author} {\bibfnamefont {E.}~\bibnamefont {Mallet}}, \bibinfo
  {author} {\bibfnamefont {P.~A.}\ \bibnamefont {Hervieux}}, \bibinfo {author}
  {\bibfnamefont {G.}~\bibnamefont {Manfredi}}, \bibinfo {author}
  {\bibfnamefont {A.}~\bibnamefont {Moreau}},\ and\ \bibinfo {author}
  {\bibfnamefont {C.}~\bibnamefont {Cirac\`{\i}}},\ }\bibfield  {title}
  {\bibinfo {title} {Influence of the electron spill-out and nonlocality on gap
  plasmons in the limit of vanishing gaps},\ }\href
  {https://doi.org/10.1103/PhysRevB.104.155435} {\bibfield  {journal} {\bibinfo
   {journal} {Phys. Rev. B}\ }\textbf {\bibinfo {volume} {104}},\ \bibinfo
  {pages} {155435} (\bibinfo {year} {2021})}\BibitemShut {NoStop}%
\bibitem [{\citenamefont {Soref}\ \emph {et~al.}(2012)\citenamefont {Soref},
  \citenamefont {Hendrickson},\ and\ \citenamefont {Cleary}}]{Soref:12}%
  \BibitemOpen
  \bibfield  {author} {\bibinfo {author} {\bibfnamefont {R.}~\bibnamefont
  {Soref}}, \bibinfo {author} {\bibfnamefont {J.}~\bibnamefont {Hendrickson}},\
  and\ \bibinfo {author} {\bibfnamefont {J.~W.}\ \bibnamefont {Cleary}},\
  }\bibfield  {title} {\bibinfo {title} {Mid- to long-wavelength infrared
  plasmonic-photonics using heavily doped n-ge/ge and n-gesn/gesn
  heterostructures},\ }\href {https://doi.org/10.1364/OE.20.003814} {\bibfield
  {journal} {\bibinfo  {journal} {Opt. Express}\ }\textbf {\bibinfo {volume}
  {20}},\ \bibinfo {pages} {3814} (\bibinfo {year} {2012})}\BibitemShut
  {NoStop}%
\bibitem [{\citenamefont {Gamal}\ \emph {et~al.}(2015)\citenamefont {Gamal},
  \citenamefont {Ismail},\ and\ \citenamefont {Swillam}}]{gamal:15}%
  \BibitemOpen
  \bibfield  {author} {\bibinfo {author} {\bibfnamefont {R.}~\bibnamefont
  {Gamal}}, \bibinfo {author} {\bibfnamefont {Y.}~\bibnamefont {Ismail}},\ and\
  \bibinfo {author} {\bibfnamefont {M.~A.}\ \bibnamefont {Swillam}},\
  }\bibfield  {title} {\bibinfo {title} {Silicon waveguides at the
  mid-infrared},\ }\href {https://doi.org/10.1109/JLT.2015.2410493} {\bibfield
  {journal} {\bibinfo  {journal} {Journal of Lightwave Technology}\ }\textbf
  {\bibinfo {volume} {33}},\ \bibinfo {pages} {3207} (\bibinfo {year}
  {2015})}\BibitemShut {NoStop}%
\bibitem [{\citenamefont {Biagioni}\ \emph {et~al.}(2015)\citenamefont
  {Biagioni}, \citenamefont {Frigerio}, \citenamefont {Samarelli},
  \citenamefont {Gallacher}, \citenamefont {Baldassarre}, \citenamefont
  {Sakat}, \citenamefont {Calandrini}, \citenamefont {Millar}, \citenamefont
  {Giliberti}, \citenamefont {Isella}, \citenamefont {Paul},\ and\
  \citenamefont {Ortolani}}]{Biagioni:15_group}%
  \BibitemOpen
  \bibfield  {author} {\bibinfo {author} {\bibfnamefont {P.}~\bibnamefont
  {Biagioni}}, \bibinfo {author} {\bibfnamefont {J.}~\bibnamefont {Frigerio}},
  \bibinfo {author} {\bibfnamefont {A.}~\bibnamefont {Samarelli}}, \bibinfo
  {author} {\bibfnamefont {K.}~\bibnamefont {Gallacher}}, \bibinfo {author}
  {\bibfnamefont {L.}~\bibnamefont {Baldassarre}}, \bibinfo {author}
  {\bibfnamefont {E.}~\bibnamefont {Sakat}}, \bibinfo {author} {\bibfnamefont
  {E.}~\bibnamefont {Calandrini}}, \bibinfo {author} {\bibfnamefont {R.~W.}\
  \bibnamefont {Millar}}, \bibinfo {author} {\bibfnamefont {V.}~\bibnamefont
  {Giliberti}}, \bibinfo {author} {\bibfnamefont {G.}~\bibnamefont {Isella}},
  \bibinfo {author} {\bibfnamefont {D.~J.}\ \bibnamefont {Paul}},\ and\
  \bibinfo {author} {\bibfnamefont {M.}~\bibnamefont {Ortolani}},\ }\bibfield
  {title} {\bibinfo {title} {{Group-IV midinfrared plasmonics}},\ }\href
  {https://doi.org/10.1117/1.JNP.9.093789} {\bibfield  {journal} {\bibinfo
  {journal} {Journal of Nanophotonics}\ }\textbf {\bibinfo {volume} {9}},\
  \bibinfo {pages} {1 } (\bibinfo {year} {2015})}\BibitemShut {NoStop}%
\bibitem [{\citenamefont {Chang}\ \emph {et~al.}(2012)\citenamefont {Chang},
  \citenamefont {Paeder}, \citenamefont {Hvozdara}, \citenamefont {Hartmann},\
  and\ \citenamefont {Herzig}}]{Chang:12}%
  \BibitemOpen
  \bibfield  {author} {\bibinfo {author} {\bibfnamefont {Y.-C.}\ \bibnamefont
  {Chang}}, \bibinfo {author} {\bibfnamefont {V.}~\bibnamefont {Paeder}},
  \bibinfo {author} {\bibfnamefont {L.}~\bibnamefont {Hvozdara}}, \bibinfo
  {author} {\bibfnamefont {J.-M.}\ \bibnamefont {Hartmann}},\ and\ \bibinfo
  {author} {\bibfnamefont {H.~P.}\ \bibnamefont {Herzig}},\ }\bibfield  {title}
  {\bibinfo {title} {Low-loss germanium strip waveguides on silicon for the
  mid-infrared},\ }\href {https://doi.org/10.1364/OL.37.002883} {\bibfield
  {journal} {\bibinfo  {journal} {Opt. Lett.}\ }\textbf {\bibinfo {volume}
  {37}},\ \bibinfo {pages} {2883} (\bibinfo {year} {2012})}\BibitemShut
  {NoStop}%
\bibitem [{\citenamefont {Ramirez}\ \emph {et~al.}(2018)\citenamefont
  {Ramirez}, \citenamefont {Liu}, \citenamefont {Vakarin}, \citenamefont
  {Frigerio}, \citenamefont {Ballabio}, \citenamefont {Roux}, \citenamefont
  {Bouville}, \citenamefont {Vivien}, \citenamefont {Isella},\ and\
  \citenamefont {Marris-Morini}}]{Ramirez:18}%
  \BibitemOpen
  \bibfield  {author} {\bibinfo {author} {\bibfnamefont {J.~M.}\ \bibnamefont
  {Ramirez}}, \bibinfo {author} {\bibfnamefont {Q.}~\bibnamefont {Liu}},
  \bibinfo {author} {\bibfnamefont {V.}~\bibnamefont {Vakarin}}, \bibinfo
  {author} {\bibfnamefont {J.}~\bibnamefont {Frigerio}}, \bibinfo {author}
  {\bibfnamefont {A.}~\bibnamefont {Ballabio}}, \bibinfo {author}
  {\bibfnamefont {X.~L.}\ \bibnamefont {Roux}}, \bibinfo {author}
  {\bibfnamefont {D.}~\bibnamefont {Bouville}}, \bibinfo {author}
  {\bibfnamefont {L.}~\bibnamefont {Vivien}}, \bibinfo {author} {\bibfnamefont
  {G.}~\bibnamefont {Isella}},\ and\ \bibinfo {author} {\bibfnamefont
  {D.}~\bibnamefont {Marris-Morini}},\ }\bibfield  {title} {\bibinfo {title}
  {Graded sige waveguides with broadband low-loss propagation in the mid
  infrared},\ }\href {https://doi.org/10.1364/OE.26.000870} {\bibfield
  {journal} {\bibinfo  {journal} {Opt. Express}\ }\textbf {\bibinfo {volume}
  {26}},\ \bibinfo {pages} {870} (\bibinfo {year} {2018})}\BibitemShut
  {NoStop}%
\bibitem [{\citenamefont {Mu}\ \emph {et~al.}(2014)\citenamefont {Mu},
  \citenamefont {Soref}, \citenamefont {Kimerling},\ and\ \citenamefont
  {Michel}}]{Mu:14}%
  \BibitemOpen
  \bibfield  {author} {\bibinfo {author} {\bibfnamefont {J.}~\bibnamefont
  {Mu}}, \bibinfo {author} {\bibfnamefont {R.}~\bibnamefont {Soref}}, \bibinfo
  {author} {\bibfnamefont {L.~C.}\ \bibnamefont {Kimerling}},\ and\ \bibinfo
  {author} {\bibfnamefont {J.}~\bibnamefont {Michel}},\ }\bibfield  {title}
  {\bibinfo {title} {Silicon-on-nitride structures for mid-infrared gap-plasmon
  waveguiding},\ }\href {https://doi.org/10.1063/1.4862795} {\bibfield
  {journal} {\bibinfo  {journal} {Applied Physics Letters}\ }\textbf {\bibinfo
  {volume} {104}},\ \bibinfo {pages} {031115} (\bibinfo {year}
  {2014})}\BibitemShut {NoStop}%
\bibitem [{\citenamefont {Gallacher}\ \emph {et~al.}(2018)\citenamefont
  {Gallacher}, \citenamefont {Millar}, \citenamefont
  {Gri\v{s}kevi\v{c}i\={u}te}, \citenamefont {Baldassarre}, \citenamefont
  {Sorel}, \citenamefont {Ortolani},\ and\ \citenamefont
  {Paul}}]{Gallacher:18}%
  \BibitemOpen
  \bibfield  {author} {\bibinfo {author} {\bibfnamefont {K.}~\bibnamefont
  {Gallacher}}, \bibinfo {author} {\bibfnamefont {R.}~\bibnamefont {Millar}},
  \bibinfo {author} {\bibfnamefont {U.}~\bibnamefont
  {Gri\v{s}kevi\v{c}i\={u}te}}, \bibinfo {author} {\bibfnamefont
  {L.}~\bibnamefont {Baldassarre}}, \bibinfo {author} {\bibfnamefont
  {M.}~\bibnamefont {Sorel}}, \bibinfo {author} {\bibfnamefont
  {M.}~\bibnamefont {Ortolani}},\ and\ \bibinfo {author} {\bibfnamefont
  {D.~J.}\ \bibnamefont {Paul}},\ }\bibfield  {title} {\bibinfo {title} {Low
  loss ge-on-si waveguides operating in the 8-14$ \mu$m atmospheric
  transmission window},\ }\href {https://doi.org/10.1364/OE.26.025667}
  {\bibfield  {journal} {\bibinfo  {journal} {Opt. Express}\ }\textbf {\bibinfo
  {volume} {26}},\ \bibinfo {pages} {25667} (\bibinfo {year}
  {2018})}\BibitemShut {NoStop}%
\bibitem [{\citenamefont {Taliercio}\ and\ \citenamefont
  {Biagioni}(2019)}]{Taliercio:2019ib}%
  \BibitemOpen
  \bibfield  {author} {\bibinfo {author} {\bibfnamefont {T.}~\bibnamefont
  {Taliercio}}\ and\ \bibinfo {author} {\bibfnamefont {P.}~\bibnamefont
  {Biagioni}},\ }\bibfield  {title} {\bibinfo {title} {Semiconductor infrared
  plasmonics},\ }\href {https://doi.org/10.1515/nanoph-2019-0077} {\bibfield
  {journal} {\bibinfo  {journal} {Nanophotonics}\ }\textbf {\bibinfo {volume}
  {8}},\ \bibinfo {pages} {949} (\bibinfo {year} {2019})}\BibitemShut {NoStop}%
\bibitem [{\citenamefont {De~Luca}\ \emph {et~al.}()\citenamefont {De~Luca},
  \citenamefont {Ortolani},\ and\ \citenamefont
  {Cirac\`{\i}}}]{deluca:epj2022}%
  \BibitemOpen
  \bibfield  {author} {\bibinfo {author} {\bibfnamefont {F.}~\bibnamefont
  {De~Luca}}, \bibinfo {author} {\bibfnamefont {M.}~\bibnamefont {Ortolani}},\
  and\ \bibinfo {author} {\bibfnamefont {C.}~\bibnamefont {Cirac\`{\i}}},\
  }\bibfield  {title} {\bibinfo {title} {Free electron harmonic generation in
  heavily doped semiconductors: the role of the materials properties},\
  }\href@noop {} {\bibinfo  {journal} {(submitted)}\ }\BibitemShut {NoStop}%
\end{thebibliography}
\end{document}